\documentclass[prb,preprint,amsmath,amssymb,superscriptaddress]{revtex4}
\usepackage{graphicx}

\begin{document}
\title{BCS-BEC crossover-like phenomena driven by quantum-size effects
in quasi-one-dimensional fermionic condensates}

\author{A. A. Shanenko}
\affiliation{TGM, Departement Fysica, Universiteit Antwerpen,
Groenenborgerlaan 171, B-2020 Antwerpen, Belgium}
\author{M. D. Croitoru}
\affiliation{Condensed Matter Theory Group, CPMOH, University of
Bordeaux I, France}
\author{A. V. Vagov}
\affiliation{Institut f\"{u}r Theoretische Physik III, Bayreuth
Universit\"{a}t, Bayreuth 95440, Germany}
\author{V. M. Axt}
\affiliation{Institut f\"{u}r Theoretische Physik III, Bayreuth
Universit\"{a}t, Bayreuth 95440, Germany}
\author{A. Perali}
\affiliation{School of Pharmacy, Physics Unit, University of Camerino,
I-62032-Camerino, Italy}
\author{F. M. Peeters}
\affiliation{TGM, Departement Fysica, Universiteit Antwerpen,
Groenenborgerlaan 171, B-2020 Antwerpen, Belgium}
\date{\today}

\maketitle
{\bf Quantum confinement is known to influence fermionic condensates,
resulting in quantum-size oscillations of superfluid/superconducting
properties. Here we show that the impact of quantum-size effects is
even more dramatic. Under realistic conditions, a significant
phase-space reconfiguration induced by quantum-size effects opens a
quasi-molecule channel in the fermionic pairing so that the
condensed pairs exhibit features typical of a molecular state. As an
illustration we consider a quasi-one-dimensional fermionic
condensate, as realized, e.g., in cigar-shaped atomic Fermi gases or
superconducting quantum wires. In this case the transverse
quantization of the particle motion favors pairing through a
coherent superposition of quantum channels that are formed due to the
grouping of single-particle levels into a series of well
distinguished subbands. Whenever the bottom of a subband approaches
the Fermi level, the longitudinal spatial distribution of fermions
in a condensed pair becomes strongly localized within the
corresponding quantum channel. The fermionic pairs in this channel
resemble molecules with bosonic character.}

In 1960's Blatt and Thompson~\cite{blatt,thom} calculated the energy
gap of an ultrathin superconducting slab as function of its
thickness and found a series of pronounced peaks. They called these
peaks {\it shape superconducting resonances}. The physics behind
this result is usually understood as follows. The spectrum
describing the electron motion in the direction perpendicular to the
film is quantized whereas a quasi-free electron motion is assumed in
the direction parallel to the film. Due to the transverse
quantization, the conduction band splits up into a series of
single-particle subbands. The lower edges (bottoms) of such subbands
are determined by the perpendicular discrete electron levels and,
so, move in energy with changing thickness $d$ (scaling as $\sim
1/d^2$). Unlike the subbands, the Fermi level exhibits a
significantly less pronounced size-dependent shift. So, the bottoms
of the subbands pass through the Fermi surface one by one when
increasing the thickness. Each time when the bottom of a subband
approaches the Fermi surface, the density of single-particle states
in the vicinity of the Fermi level increases, resulting in an
enhancement of basic superconducting quantities, e.g., the critical
temperature, the order parameter and the excitation gap. Thus, the
formation of single-particle subbands due to quantum confinement
manifests itself through {\it quantum-size oscillations} of the
superconducting properties driven by a series of shape
superconducting resonances (below they are also referred to as
quantum-size superconducting/superfluid resonances or size-dependent
superconducting/superfluid resonances).

For several decades after the paper by Blatt and Thompson, only few
experimental groups reported possible signatures of quantum-size
oscillations in superconducting films~\cite{komnik,orr,pfenn}.
However, poor sample purity and significantly limited size control
prevented definite conclusions. Blatt and Thompson also argued (see
Refs.~\onlinecite{thom1,thom2,kenn}) that similar physics could be
expected for nucleon pairing in atomic nuclei as a consequence of
the shell structure of the single-particle spectrum. This expectation
was found to be in agreement with experimental data~\cite{nem} and,
for the next 40 years, atomic nuclei were the only system where
quantum-size effects on the BCS fermionic pairing were investigated
both experimentally and theoretically (for more details, see
Ref.~\onlinecite{hil} and references therein). Advances in
microstructuring of superconductors in the 90's renewed interest in
quantum-size effects in superconducting systems. The work of Blatt
and Thompson was extended by Bianconi, Perali and coworkers
\cite{Perali96,Bianconi98} to quantum striped superconductors, where
a sizeable amplification of the critical temperature and excitation
gap was found in the presence of a shape superconducting resonance.
Recent developments in nanofabrication have resulted in high-quality
superconducting nanosystems, and the quantum-size oscillations of
the critical superconducting temperature $T_c$ were eventually
observed in single-crystalline atomically uniform in thickness ${\rm
Pb}$ nanofilms\cite{guo,eom,qin}. Furthermore, the results of a
numerical self-consistent solution of the Bogoliubov-de Gennes (BdG)
equations in Ref.~\onlinecite{shan} showed that the quantum-size
superconducting resonances were responsible for a systematic
thickness-dependent shift-up of $T_c$ found in high-quality aluminum
and tin superconducting nanowires~\cite{tian,tombr,zgi,alt}. Very recently
quantum-size oscillations of the excitation gap were also reported
for tin superconducting nanograins~\cite{bose}.

Cooling of trapped fermionic atoms down to ultra-low temperatures
such that the atomic gas is Fermi degenerate (see, e.g.,
Ref.~\onlinecite{bloch}), resulted in another system promising for
the investigation of quantum-size effects in fermionic condensates.
In the case of trapped atomic Fermi gases the single-particle levels
can be tuned by laser light or magnetic fields through, e.g.,
changing the spatial dimensions of the trap. In particular, for a
pancake- or cigar-shaped geometry the perpendicular trapping
frequency $\omega_{\perp}$ is much larger than the parallel one
$\omega_{||}$. In this case single-particle states form well
distinguished subbands and the physical picture turns out to be
similar to that in superconducting metallic nanowires and nanofilms,
i.e., a quantum-size superfluid resonance can be expected when the
bottom of a single-particle subband is located in the vicinity of
the Fermi level. In particular, quantum-size oscillations of the
critical temperature and order parameter were recently calculated
for a superfluid Fermi gas confined in a quasi-two-dimensional
(quasi-2D) harmonic trap~\cite{torma}. We note that fermionic
condensates are now attainable even in the tight-confinement regime
with only one perpendicular single-particle level below the Fermi
energy\cite{bloch}. Recently, experimental results on the first
indications on quantum-size oscillations of the cloud-size aspect
ratio in a harmonically trapped quasi-2D Fermi gas became
available~\cite{vale}. Maintaining the atomic cloud at the lowest
attainable temperature and at a low number of atoms, the group of
Vale~\cite{vale} was able to detect the filling of the individual
perpendicular levels, following in a quantitative way the
dimensional crossover from 2D to 3D in a $^6Li$ Fermi gas, while
keeping the 3D character of the two-particle scattering. This work
clearly demonstrates that the experimental conditions to explore
such a dimensional crossover and the accompanying quantum-size
effects are presently achievable with ultracold atoms.

In this paper, we consider a quasi-1D fermionic condensate and show
that the impact of the formation of single-particle subbands is more
dramatic than simply resulting in oscillations of $T_c$ or the
excitation energy gap. Significant reconfiguration of the phase
space due to grouping of single-particle levels into a series of
subbands (here quasi-1D means that we deal with more than one 1D
subband of single-particle states) results in opening of a
quasi-molecule channel in the fermionic pairing each time when the
bottom of a single-particle subband approaches the Fermi surface.
Due to a "depletion" of the longitudinal Fermi motion in such a
subband the longitudinal spatial distribution of fermions inside a
condensed fermionic pair squeezes and becomes strongly localized. In
this case the system behaves similar to the well-known crossover
from the Bardeen-Cooper-Schrieffer (BCS) pairs to the Bose-Einstein
condensation (BEC) of quasi-molecules driven through the Feshbach
resonance in ultracold Fermi gases~(see, e.g.,
Ref.~\onlinecite{bloch}). However, very different from previous
case, the effect of interest is now controlled by the quantization
of the perpendicular particle motion. By changing the transverse
size of the system, one can change the energetic positions of the
transverse single-particle discrete levels (bottoms of the
single-particle subbands) with respect to the chemical potential
$\mu$. This leads to a significant redistribution of the kinetic
energy between the transverse and longitudinal degrees of freedom in
a subband and, so, to the above mentioned "depletion" of the
longitudinal Fermi motion when the bottom of this subband approaches
$\mu$. Below, for illustration, we investigate a cigar-shaped atomic
Fermi gas and a superconducting quantum wire. Similar results are
also expected for quasi-2D fermionic condensates.

\section{Harmonically trapped quasi-1D superfluid Fermi gas}

We begin with the BCS self-consistent equation for the spin-singlet
$s$-wave pairing gap $\Delta_{\bf k}$ of a $3D$ Fermi gas with an
attractive interaction, i.e.,
\begin{equation}
\Delta_{{\bf k}}=-\sum\limits_{{\bf k}'} V_{{\bf k}{\bf
k}'}\,\Delta_{{\bf k}'}\,\tanh(\beta E_{{\bf k}'}/2)\,
\biggl[\frac{1}{2E_{{\bf k}'}} - \frac{1}{2T_{{\bf k}'}}\biggr],
\label{3D}
\end{equation}
where $T_{\bf k}=\hbar^2k^2/2M$ is the single-particle dispersion,
$E_{\bf k} =\sqrt{(T_{\bf k} -\mu)^2 + \Delta_{\bf k}^2}$ stands for
the quasiparticle spectrum, $\mu$ is the chemical potential and
$\beta$ is the inverse temperature ($\beta=1/k_BT$). The standard
scattering length regularization is introduced in the gap equation
(1) to avoid ultraviolet divergency in 3D. The interaction matrix
element in Eq.~(\ref{3D}) can be written in the form
\begin{equation}
V_{{\bf k}{\bf k}'}=-g\int {\rm d}^3r |\varphi_{\bf k}({\bf
r})|^2|\varphi_{{\bf k}'}({\bf r})|^2, \label{vkk}
\end{equation}
where $g=4\pi\hbar^2 |a|/M$, with $a < 0$ the 3D $s$-wave scattering
length, and $\varphi_{\bf k}({\bf r})$ is the single-particle wave
function that is proportional to $e^{\imath{\bf k}{\bf r}}$~(plane
waves) in $3D$.

When switching to a Fermi gas confined in a trap, the particle
momentum label in Eqs.~(\ref{3D}) and (\ref{vkk}) should be replaced
by an index for discrete energy levels because the relevant wave
functions are not plane waves but solutions of the one-particle
Schr\"odinger equation for the corresponding confined geometry. In
this section we consider harmonically trapped fermions in an axially
symmetric confining potential $M(\omega_{\perp}^2 \rho^2 +
\omega_{||}^2 \,z^2)/2$~(with $\omega_{\perp} \gg \omega_{||}$),
where cylindrical coordinates are invoked. Due to the axial symmetry
the relevant quantum numbers can be chosen as $n=0,1,\ldots$, the
radial quantum number; $m=0, \pm1, \pm2, \ldots$, the azimuthal
quantum number; and $j=0,1,\ldots$, the quantum number associated
with the parallel (longitudinal) motion. For the aspect ratio of the
trap we have $l_{||}/l_{\perp}\gg 1$~(with
$l_{||}=\sqrt{\hbar/M\omega_{||}}$ and $l_{\perp}= \sqrt{\hbar
/M\omega_{\perp}}$) and, so, using the 3D pseudopotential
$g=4\pi\hbar^2 |a|/M$ seems problematic due to the possible 1D
character of the two-particle scattering. However, the aspect ratio
is not actually the quantity that controls the character of
scattering in a cigar-shaped many-particle system. The analysis of
the two-particle scattering in an axially confined geometry, see,
e.g., Ref.~\onlinecite{petrov}, shows that the scattering rapidly
becomes three dimensional when the number of contributing subbands
(perpendicular levels) is more than 1. For parameters considered in
this work this number varies from $6$ to $> 20$. Thus we are far
beyond the regime of the effectively 1D scattering and, so, any
issue related to the confinement-induced Feshbach resonance in a 1D
wave guide and the formation of confinement-induced molecules in a
1D Fermi gas (see, e.g., Refs.~\onlinecite{bloch} and
\onlinecite{bergeman,moritz,tokatly,fuchs}) are beyond the scope of
the present work. To avoid any confusion, we would like to note once
again that quasi-1D in the present work means that more than one
transverse single-particle level contribute to the basic physical
quantities.

By numerically solving Eqs.~(\ref{3D}) (with ${\bf k} \to \nu
=\{n,m,j\}$), we self-consistently calculate for a given temperature
a set of pairing gaps $\Delta_{\nu}$. Then, the critical temperature
$T_c$ can be found as the temperature above which only the trivial
($\Delta_{\nu}=0$) solution to the problem exists. To obtain more
information, e.g., concerning the spatial distribution of the pair
condensate and the fermionic pairing correlations, we should take
into account that Eq.~(\ref{3D}) follows from the Bogoliubov-de
Gennes equations provided that the particle-like and hole-like wave
functions $u_{\nu}({\bf r})$ and $v_{\nu}({\bf r})$ are approximated
as (see, e.g., Refs.~\onlinecite{shan1,croit})
\begin{equation}
u_{\nu}({\bf r})={\cal U}_{\nu}\varphi_{\nu}({\bf r}),\quad v_{\nu}
({\bf r})= {\cal V}_{\nu}\varphi_{\nu}({\bf r}),
\label{ander}
\end{equation}
with (${\cal U}_{\nu}$ and ${\cal V}_{\nu}$ are taken real)
$$
{\cal U}^2_{\nu}=\frac{1}{2}\Bigl(1+\frac{T_{\nu}-\mu}{E_{\nu}}
\Bigr),\quad {\cal V}^2_{\nu}=\frac{1}{2}\Bigl(1-\frac{T_{\nu}-
\mu}{E_{\nu}}\Bigr),
$$
and $\varphi_{\nu}({\bf r})=\vartheta_{nm}(\rho,\varphi) \chi_j(z)$,
where $\vartheta_{nm}(\rho,\varphi)$ and $\chi_j (z)$ are the
eigenfunctions of the 2D (iso\-tropic) and 1D harmonic oscillators,
respectively. This approximation accounts for the pairing of time
reversed states, and, as known since the pioneering paper by
Anderson\cite{and}, such a simplification yields quite reasonable
results in the presence of time reversal symmetry. Then, based on
the Bogoliubov transformation of the field operators
$\hat{\psi}_{\uparrow}({\bf r})$ and $\hat{\psi}_{\downarrow}({\bf
r})$ to the quasiparticle quantum amplitudes $\gamma_{\nu,
\uparrow}$ and $\gamma_{\nu,\downarrow}$, i.e.,
$$
\hat{\psi}_{\uparrow}({\bf r})= \sum_{\nu}\Bigl[u_{\nu}({\bf r})
\gamma_{\nu,\uparrow} - v^*_{\nu}({\bf r})\gamma^{\dagger}_{
\nu,\downarrow}\Bigr], \quad\hat{\psi}_{\downarrow}({\bf r})= \sum_{
\nu}\Bigl[u_{\nu}({\bf r})\gamma_{\nu,\downarrow} + v^*_{\nu} ({\bf
r})\gamma^{\dagger}_{\nu,\uparrow}\Bigr],
$$
one can study the superfluid pair correlations and the spatial
distribution of the pair condensate. In particular, introducing the
Cooper-pair ``wave function" $\Psi({\bf r},{\bf r}')=
\langle\hat{\psi}_{\uparrow}({\bf r})\hat{\psi}_{\downarrow} ({\bf
r})\rangle$, we find
\begin{equation}
\Psi({\bf r},{\bf r}')=\sum\limits_{nm} \Psi_{nm}({\bf r}, {\bf
r}'),\quad \Psi_{nm}({\bf r},{\bf r}')=\vartheta_{nm}
(\rho,\varphi)\,\vartheta^*_{nm}(\rho',\varphi')\;\psi_{nm}
(z,z'),\label{Psi}
\end{equation}
with
\begin{equation}
\psi_{nm}(z,z')=\sum\limits_j\,\chi_j(z)\;\chi^*_j(z')
\;\Delta_{nmj}\;\tanh\Bigl(\beta E_{nmj}/2\Bigr)
\,\biggl[\frac{1}{2E_{nmj}}-\frac{1}{2T_{nmj}}\biggr], \label{psiz}
\end{equation}
where $T_{nmj}=\hbar\omega_{\perp}(1 + 2n + |m|) + \hbar
\omega_{||}(j+1/2)$. It is worth noting that the regularization term
$1/(2T_{nmj})$ appears in the parenthesis of Eq.~(\ref{psiz}) as a
simple extension to spatially nonuniform systems of the
regularization used in Eq.~(\ref{3D}). It is important to note that
we have verified that different possible regularization schemes lead
to insignificant differences in the physical quantities, preserving
the basic conclusions of the present work, see a more detailed
discussion in the end of this section. The single-particle states
with the same $n$ and $m$ form a quasi-continuum. Therefore, it is
natural to introduce single-particle subbands denoted by
$(n,m)$~(see Fig. 1 in Supplementary Information) and the subband
dependent fermionic-pair "wave function" defined by Eq.~(\ref{Psi}).
We remark that treating $\Psi({\bf r},{\bf r}')$ as the wave
function of a condensed pair of fermions goes back to the classical
papers by Gor'kov\cite{gor} and Bogoliubov\cite{bog} and is directly
related to the conventional interpretation of the order parameter
$\Delta ({\bf r})=g\Psi({\bf r},{\bf r})$ as the center-of-mass
Cooper-pair wave function. However, there exist also other ways to
introduce the fermionic-pair wave function (see, e.g.,
Ref.~\onlinecite{duk} and \onlinecite{ket}). Nevertheless, all these
variants were shown\cite{duk} to result in the bulk Cooper-pair
radius being proportional to the ratio of the Fermi velocity to the
energy gap. In the present paper we are interested in the
longitudinal Cooper-pair size in a quasi-1D fermionic condensate
which can be calculated as
\begin{equation}
\xi_0=\left[\frac{\langle\Psi| (z-z')^2|\Psi\rangle}{\langle\Psi|
\Psi\rangle}\right]^{1/2}, \label{xi0}
\end{equation}
with $\Psi$ given by Eq.~(\ref{Psi}). Below we deal also with the
subband-dependent longitudinal size of condensed fermionic pairs
$\xi^{(nm)}_0$ which is controlled by $\psi_{nm}(z,z')$ and
calculated from Eq.~(\ref{xi0}) with $\Psi({\bf r},{\bf r}')$
replaced by $\psi_{nm}(z, z')$, i.e., $\xi^{(nm)}_0=[\langle
\psi_{nm}|(z-z')^2|\psi_{nm} \rangle/\langle \psi_{nm}|
\psi_{nm}\rangle]^{1/2}$.

\begin{figure*}
\includegraphics[width=0.7\linewidth]{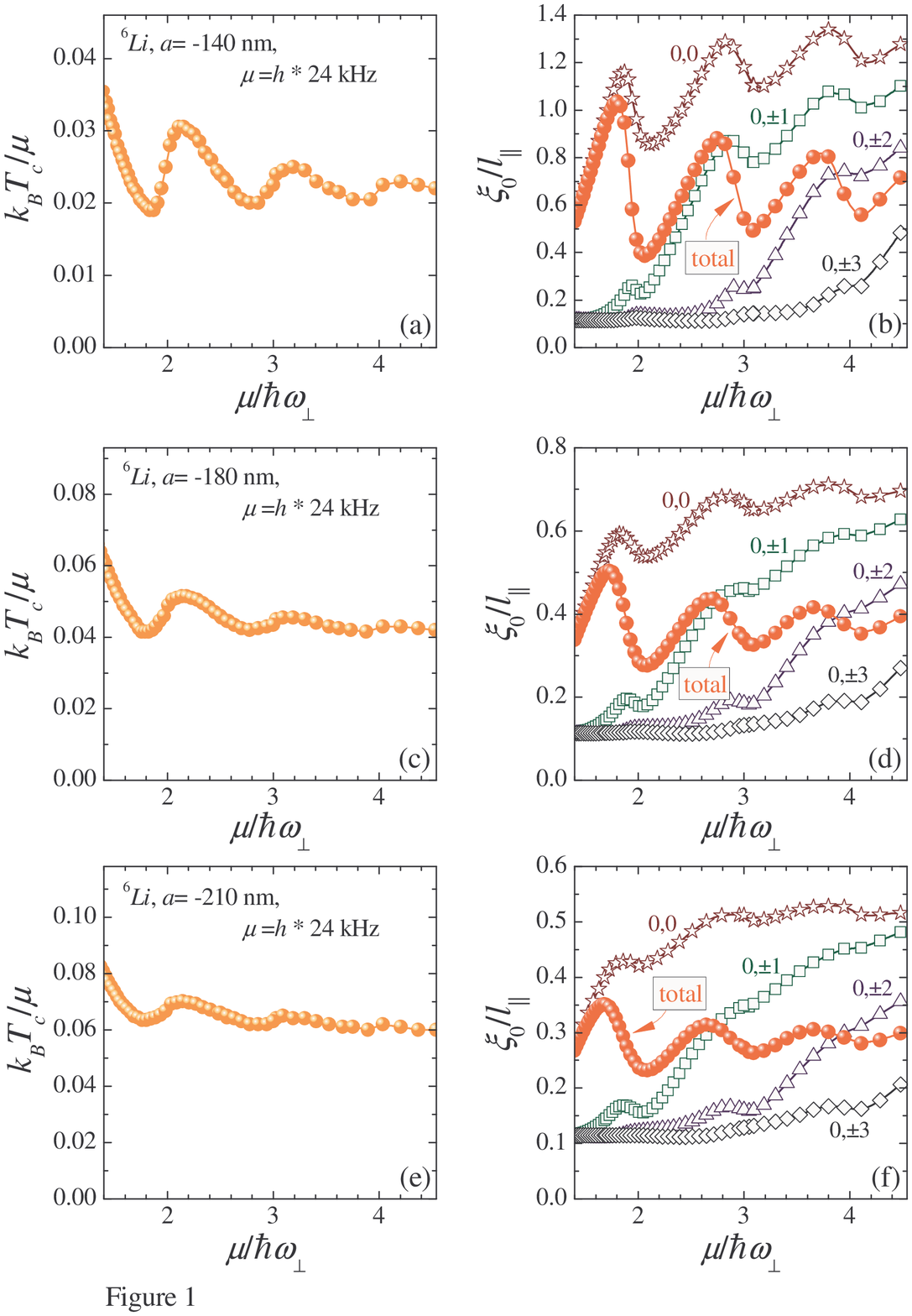}
\caption{Superfluid Fermi gas confined in a cigar-shaped harmonic
trap: the critical temperature $T_c$~(in units of $\mu$) and the BCS
coherence length $\xi_0$~(in units of $l_{||}$) versus
$\mu/\hbar\omega_{\perp}$. Panels (a) and (b) for $a=-140\,{\rm
nm}$; (c),(d) for $a=-180\,{\rm nm}$; (e) and (f) for $a=-210\, {\rm
nm}$. The subband-dependent coherence lengths $\xi^{(nm)}_0$ for
$(n,m)=(0,0),\,(0,\pm1)\,(0,\pm2)$ and $(0,\pm3)$ are also given to
compare with $\xi_0$.} \label{fig1}
\end{figure*}

For a numerical solution of Eq.~(\ref{3D}) (with ${\bf k} \to \nu=
\{n,m,j\}$), we consider a quasi-1D harmonically trapped mixture of
$^6{\rm Li}$ atoms in the two lowest spin states, $|F,m_F\rangle =
|1/2,1/2\rangle$ and $|1/2,-1/2\rangle$. The pair interaction
between these states can be significantly modified by an external
magnetic field by means of the formation of a broad Feshbach
resonance\cite{sol}. In particular, typical values of the $s$-wave
scattering length $a$ at the BCS side of this Feshbach resonance
varies\cite{sol} from $-250$ to $-100\,{\rm nm}$ dependent on the
magnetic field $B$ (theoretically $a$ goes to $-\infty$ when
approaching the point of the Feshbach resonance $B=0.83\,{\rm kG}$).
In Fig.~\ref{fig1} our numerical results are shown for $a=-140$,
$-180$ and $-210\,{\rm nm}$~[panels (a,b), (c,d) and (e,f),
respectively]. The chemical potential and the longitudinal frequency
are kept constant, i.e., $\mu=h \cdot 24\,{\rm kHz}$ and
$\hbar\omega_{||} = 0.01\mu$ ($\omega_{||}/2\pi = 240\,{\rm Hz})$,
while the trapping frequency in the perpendicular direction varies
in such a way that discrete single-particle levels for the
perpendicular single-particle motion pass through the Fermi surface
when the ratio $\mu/\hbar\omega_{\perp}$ reaches $2,\,3, \,4$ etc.
In our cigar-shaped confining geometry $\omega_{||}$ is taken much
smaller than $\omega_{\perp}\sim \mu/\hbar$. The particular value of
$\omega_{||}$ in this regime appears to be of no importance (changes
less than a few percent were found for $\omega_{||}=2\pi\cdot300
{\rm Hz}$ and $2\pi\cdot 480{\rm Hz}$). The chemical potential is
chosen such that experimentally accessible values of the particle
density and the ratio $T_c/T_F$ are obtained, with $T_F$ denoting
the Fermi temperature (for more details, see the discussion at the
end of this section).

Figure~\ref{fig1} (a) shows our numerical results for $T_c$~(in
units of $\mu/k_B$) as a function of $\mu/\hbar\omega_{\perp}$
calculated for $a=-140\,{\rm nm}$. As seen, each time when a
discrete transverse level crosses $\mu$, $T_c$ is enhanced, i.e., a
superfluid quantum-size resonance occurs. As a result, $T_c$
oscillates with changing $\omega_{\perp}$. These oscillations are
damped: their amplitude is reduced when $\omega_{\perp}$ decreases
($l_{\perp}$ increases). Such a decay is typical of quantum-size
oscillations of the basic physical quantities in quasi-1D and
quasi-2D fermionic condensates\cite{eom,qin,shan,torma}. The reason
is that the number of single-particle subbands making a contribution
increases with $l_{\perp}$. In particular, for
$\mu/\hbar\omega_{\perp}=2$ there are six subbands that are
responsible for the formation of $99\%$ of the condensate [subbands
$(n,m)=(0,0), \,(0,\pm1),\,(0,\pm2)$ and $(1,0)$, major contribution
is due to $(n,m)=(0,\pm1)$]. For $\mu/\hbar\omega_{\perp}=4$ there
are already $21$ contributing subbands. The larger this number is,
the less pronounced is the increase in the relevant density of
states when a new perpendicular level crosses $\mu$. Then,
quantum-size resonances are weakened and the corresponding
oscillations are finally washed out. Another important feature of
such oscillations is that their amplitude is also reduced when the
pair interaction is enhanced (i.e., $|a|$ increases). A stronger
pair interaction results in larger pairing gaps. In this case
contributions of the levels even far below or above $\mu$ become of
importance. As a result, passages of perpendicular levels through
the Fermi surface have less pronounced and rather smoothed effects.
This is illustrated by Figs.~\ref{fig1} (c) and (e), where
$k_BT_c/\mu$ is given versus $\mu/\hbar\omega_{\perp}$ for
$a=-180\,{\rm nm}$ and $a=-210\,{\rm nm}$. Note that in the present
work $T_c$ is evaluated at the mean-field level. On the other hand,
it is well known that the mean field critical temperature through
the BCS-BEC crossover (driven by a change in the coupling $k_Fa$)
corresponds to a characteristic temperature of the pair formation
and pseudogap opening, while the critical temperature for the
superfluid transition is renormalized by pair fluctuations. In the
simplest approach, the renormalized critical temperature can be
evaluated by using a non self-consistent t-matrix approach, as shown
in Refs.~\onlinecite{Perali02} and \onlinecite{Perali04}. For our
choice of parameters we have $-1.4 < (k_Fa)^{-1}< -0.9$ (with $k_F$
the Fermi wavevector in the center of the trap, see the discussion
at the end of this section). These values of $k_Fa$ are between the
BCS and the crossover regimes, where the suppression of the critical
temperature due to the pair fluctuations is not larger than $16\%$,
thanks also to the fact that in trapped systems the fluctuation
region shrinks with respect to the homogeneous case. Therefore, the
amplitude and period of the oscillations of $T_c$ induced by quantum
confinement as shown in Fig.~\ref{fig1}, will not be importantly
modified by fluctuations.

Based on the above results for $T_c$, one can also anticipate
similar oscillations of the longitudinal BCS coherence length
$\xi_0$. These oscillations are shown (red solid circles) in
Figs.~\ref{fig1}(b), (d) and (f) for $a=-140$, $-180$, and
$-210\,{\rm nm}$, respectively. Naive expectation based on the
ordinary BCS picture suggests that $\xi_0 \propto \hbar
v_F/(k_BT_c)$, with the Fermi velocity $v_F =\hbar k_F/m_e$.
However, this expectation does not match our numerical results. As
seen from Fig.~\ref{fig1}(a), the coherence length $\xi_0$ drops by
a factor of $2.7$  when $\mu/\hbar\omega_{\perp}$ increases from
$1.8$ to $2.1$. While according to the above naive estimation,
$\xi_0$ is expected to fall only by about $40\%$~(the critical
temperature increases by about $50\%$ with almost negligible change
of $v_F$). Similar difference can be found between panels (d) and
(f). The reason for this difference is the formation of the
multi-subband structure. For illustrative purposes, the
subband-dependent coherence length $\xi^{(nm)}_0$ is also shown in
Figs.~\ref{fig1}(b), (d) and (f) for $(n,m)=(0,0),\,
(0,\pm1),\,(0,\pm2)$ and $(0,\pm3)$. At $\mu=2\hbar \omega_{\perp}$
the bottoms of the two degenerate subbands with $(n,m)=(0,\pm1)$
pass through the Fermi surface and the corresponding quantum-size
resonance develops (a subband whose bottom is in the vicinity of the
Fermi surface is referred to as resonant). At this point subbands
$(0,\pm1)$ make the major contribution to the basic physical
quantities. In particular, their common contribution to the integral
of $|\Psi({\bf r}_1, {\bf r}_2)|^2$ is about $70$ - $75\%$. In other
words, we may state that almost $70$ -$75\%$ of the fermionic pairs
come from these subbands. As a result, being close to
$\xi^{(0,0)}_0$ at $\mu < 1.8\,\hbar \omega_{\perp}$, $\xi_0$
changes its trend abruptly and approaches
$\xi^{(0,1)}_0=\xi^{(0,-1)}_0$ when the perpendicular level $2\hbar
\omega_{\perp}$ crosses $\mu$. For $a=-140\,{\rm nm}$, panel (b),
this change is more dramatic because of a more pronounced difference
between $\xi^{(0,0)}_0$ and $\xi^{(0,\pm1)}_0$. When $|a|$
increases, $\xi^{(0,0)}_0$ and $\xi^{(0,\pm1)}_0$ become closer to
one another and, as a result, the effect is weakened. A similar
weakening also occurs for upper-level resonances, i.e., associated
with the perpendicular levels $3\hbar\omega_{\perp}$,
$4\hbar\omega_{\perp}$, etc. This is mostly because the
inter-subband energy spacing is reduced ($\hbar\omega_{\perp}$
decreases) so that the difference between the subband-dependent
lengths becomes less and less pronounced. It is worth noting that
the above naive estimation of the longitudinal Cooper-pair size,
i.e., $\propto \hbar v_F/(k_BT_c)$, is not totally irrelevant: it
yields reasonable results for $\xi^{(nm)}_0$ in a subband with the
bottom far below $\mu$. For instance, from Fig.~\ref{fig1}(b) one
finds that $\xi^{(0,0)}_0$ reduces by about $30\%$ when $\mu/\hbar
\omega_{\perp}$ increases from $1.8$ to $2.1$, which is close to the
drop of $40\%$ following from the above simplified estimation. In
addition, one can compare $\xi^{(0,0)}_0$ at
$\mu/\hbar\omega_{\perp}=1.8$~(below the resonance point
$\mu/\hbar\omega_{\perp}=2.0$) for different panels (different
interaction strength) in Fig.~\ref{fig1}. As seen, this quantity
scales approximately as the inverse critical temperature, which
agrees with the estimate $\hbar v_F/(k_BT_c)$. Now, let us check
what happens with a subband whose bottom is in the vicinity of
$\mu$. Here the naive estimation of the longitudinal Cooper-pair
size is not longer in agreement with our numerical results. In
particular, $\xi^{(0,\pm1)}_0$ taken at $\mu = 2\hbar\omega_{\perp}$
does not scale as $1/T_c$ when changing $a$~(the bottoms of the
subbands with $(n,m)=(0,\pm1)$ crosses the Fermi level at $\mu =
2\hbar \omega_{\perp}$). Here a reasonable agreement can be achieved
when assuming a less sensitive scaling, e.g., $1/\sqrt{T_c}$. The
same occurs for  $\xi^{(0,\pm2)}_0$ and $\xi^{(0,\pm3)}_0$ at $\mu =
3\hbar \omega_{\perp}$ and $4\hbar\omega_{\perp}$, respectively.
However, scaling $1/\sqrt{T_c}$ does not match at all when the
bottom of a subband goes above the Fermi surface. As seen from
panels (b), (d) and (f), $\xi_{nm}$'s for $(n,m)=(0,\pm1)$,
$(0,\pm2)$ and $(0,\pm3)$ are close to one another for $\mu <
2\hbar\omega_{\perp}$ and practically do not change with $a$. It is
worth noting that $\xi^{(1,0)}_0$ and $\xi^{(0,\pm2)}_0$ are not
exactly the same in spite of the fact that subbands $(0,\pm2)$ and
$(1,0)$ are degenerate. This difference is within several percent
and appears due to a difference in the relevant interaction matrix
elements, which results in $\psi_{1,0}(\rho,\varphi) \not=
\psi_{0,\pm2} (\rho,\varphi)$. A similar difference appears between
$ \xi^{(0,\pm3)}_0$  and $\xi^{(1,\pm1)}_0$.

\begin{figure}
\includegraphics[width=0.85\linewidth]{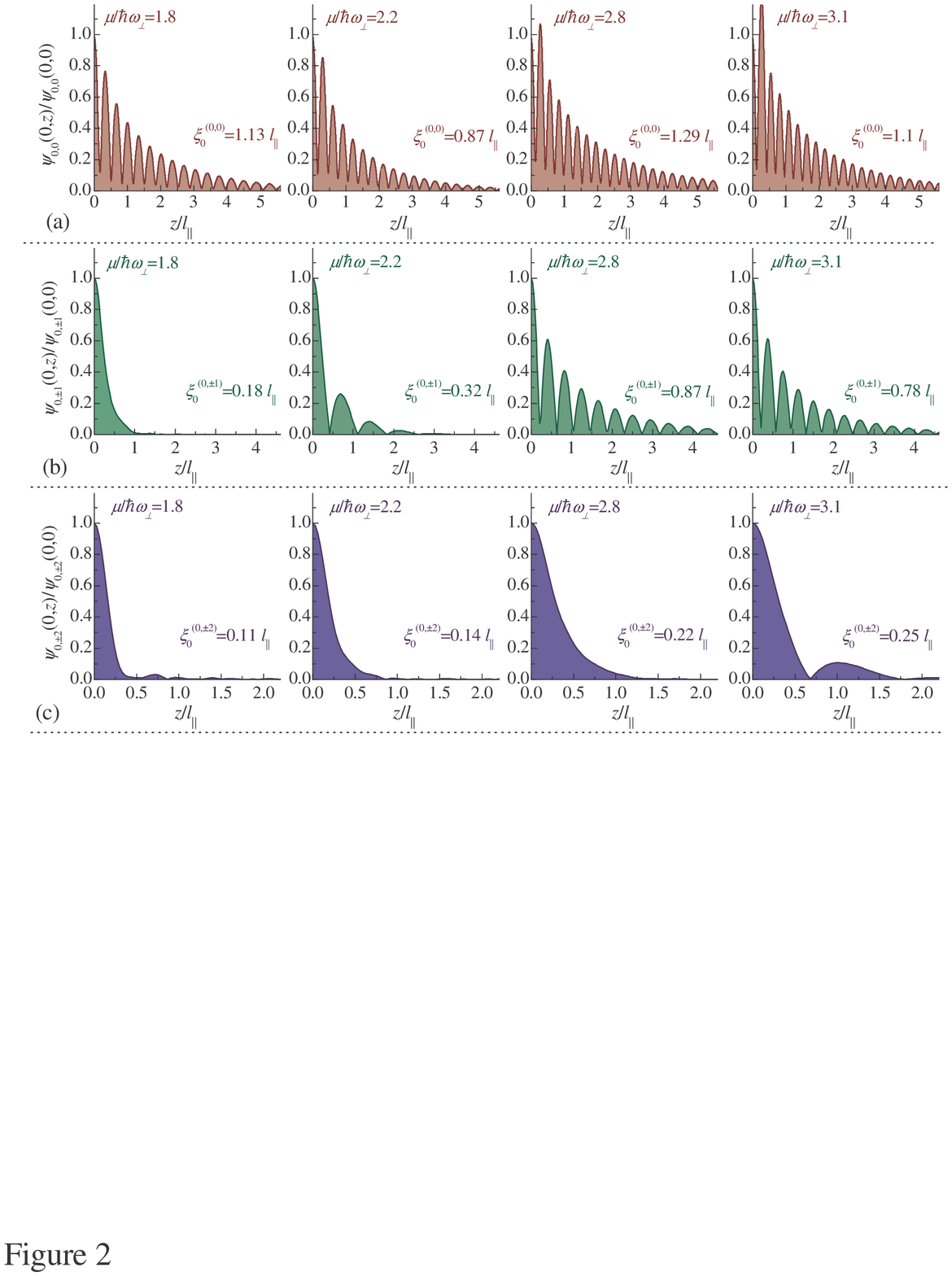}
\caption{Cigar-shaped superfluid Fermi gas: the absolute value of the subband-dependent longitudinal fermionic pair ``wave function"
$\psi_{nm}(0,z)/\psi(0,0)$ (i.e., one fermion is fixed at the origin
and another is at the point $z$) versus $z/l_{||}$ for $a=-140\,{\rm
nm}$ at $\mu/\hbar\omega_{\perp}=1.8,\,2.2,\,2.8$ and $3.1$.
Panels (a), (b) and (c) show the results for $(n,m)=(0,0),\,
(0,\pm1)$ and $(0,\pm2)$, respectively.} \label{fig2}
\end{figure}

To go in more detail about the subband-dependent fermionic pairing,
we consider how $\psi_{nm}(z,z')$ decays with increasing $|z-z'|$
(the characteristic length for this decay is $\xi^{(nm)}_0$) for
different energetic positions of the bottom of the corresponding
single-particle subband with respect to $\mu$.
Figures~\ref{fig2}(a), (b) and (c) show $\psi_{nm}(0,z)$~[given in
units of $\psi_{nm}(0,0)$] as a function of $z$ for $(n,m) =
(0,0),\,(0,\pm1)$ and $(0,\pm2)$, respectively (here $a=-140\,{\rm
nm}$). Curves for $\mu/\hbar \omega_{\perp} = 1.8,\,2.2,\,2.8$ and
$3.1$ are given in each panel. For all these values of
$\mu/\hbar\omega_{\perp}$ the bottom of single-particle subband
$(0,0)$~[see panel (a)] is situated far below the Fermi surface and,
as a result, $\psi_{0,0}(0,z)$ exhibits features typical for loosely
bound Cooper pairs (see, e.g., Ref.~\onlinecite{duk}), i.e., we
observe an exponentially decaying curve with superimposed fast
oscillations with period $2\pi/k^{(0,0)}_F$, where $k^{(n,m)}_F=
\sqrt{\frac{2M}{\hbar^2} \,[\mu -\hbar \omega_{\perp} (1+2n+|m|)]}$
is the subband-dependent longitudinal Fermi wavevector. Unlike
Fig.~\ref{fig2}(a), Figs.~\ref{fig2}(b) and (c) show results that
significantly differ from the typical BCS behavior [except of
$\mu/\hbar \omega_{\perp}= 2.8$ and $3.1$ in panel (b), where the
bottoms of the subbands with $(n,m)=(0,\pm1)$ goes below the Fermi
surface]. These results resemble the behavior of a wave function of
a real bound state in a two-body problem unless the bottom of the
corresponding subband goes below the Fermi level, see
$\mu/\hbar\omega_{\perp}=1.8$ for $(n,m)=(0,\pm1)$ in panel (b) and
$\mu/\hbar\omega_{\perp}= 1.8,\,2.2$ and $2.8$ for $(n,m)=(0,\pm2)$
in panel (c). We note that for the particle densities in the center
of the trap $n_p \approx 2 \times 10^{12}$ - $5 \times 10^{12}\,{\rm
cm}^{-3}$~(this corresponds to our choice of the chemical potential,
see details below) the mean inter-particle distance $1/n^{1/3}_p$
varies in the range $0.6$ - $0.8 \,{\rm \mu m}$. As $l_{||}\approx
2.5 {\rm \mu m}$, than $\xi^{(0,\pm1)}_0$ is less than the mean
inter-particle distance for $\mu/\hbar \omega_{\perp}=1.8$~[see
Fig.~\ref{fig2}(b)], and the same is true for $\xi^{(0,\pm2)}_0$
when $\mu/\hbar\omega_{\perp} \leq 2.8$~[see Fig.~\ref{fig2}(c)].
Hence, we observe a universal behavior (not depending on particular
quantum numbers) of $\psi_{nm} (0,z)$ that exhibits a strong
localization when the perpendicular level
$\hbar\omega_{\perp}(1+2n+|m|)$ crosses the Fermi surface. Such a
localization of the pair distribution is to a great extent similar
to the behavior of the fermionic pair-wave function at the ordinary
BCS-BEC crossover driven by changing the pair interaction
strength~\cite{duk}. What is the reason for the longitudinal
localization of the pair distribution in our case? When the bottom
of a single-particle subband is situated far below the Fermi level,
the ratio of the interaction energy to the longitudinal kinetic
energy in such a subband is small. As a result, we obtain the
ordinary BCS picture for the longitudinal distribution of fermions
in a condensed fermionic pair. However, when the bottom approaches
the Fermi level, the longitudinal Fermi motion is depleted,
resulting in a significant decrease of the ratio of the longitudinal
kinetic energy to the interaction energy and, as a consequence, in
the longitudinal squeezing of a condensed fermionic pair. The
depletion of the longitudinal Fermi motion in a subband with the
bottom close to $\mu$ can be seen from Figs.~\ref{fig2}(b): the
period of oscillations in $\psi_{0,\pm1}(0,z)$ decreases when
passing from $\mu=2.2 \hbar\omega_{\perp}$ to $3.1\hbar
\omega_{\perp}$. Here the question can arise if a subband with the
bottom above the Fermi level can contribute to the pair condensate.
For an ideal Fermi gas such a subband is not populated. However,
this is not the case in the presence of superfluid/superconducting
correlations that smoothen the Fermi surface. If the energy spacing
between the bottom of a subband and the Fermi surface is about or
less than the relevant pairing gap, this subband still contributes.
In the opposite case its role is diminished and becomes negligible
with an increase of the above spacing. For instance, as seen from
Figs.~\ref{fig1}(b), (d) and (f) the subbands with $(n,m)=(0,\pm1)$
make a significant contribution to the coherent properties even at
$\mu =1.9\hbar\omega_{\perp}$. However, their contribution becomes
negligible for $\mu < 1.8\hbar\omega_{\perp}$, where the total
longitudinal length $\xi_0$ approaches $\xi^{(0,0)}_0$. It is also
of importance to note that the ratio of the interaction energy to
the total kinetic energy is not affected by the perpendicular
quantization, i.e., it remains small enough even when the bottom of
one of the single-particle subbands approaches the Fermi surface.
The effect of interest is due to a redistribution of the kinetic
energy between the parallel and perpendicular degrees of freedom in
the subband whose bottom approaches $\mu$.

\begin{figure}
\includegraphics[width=0.85\linewidth]{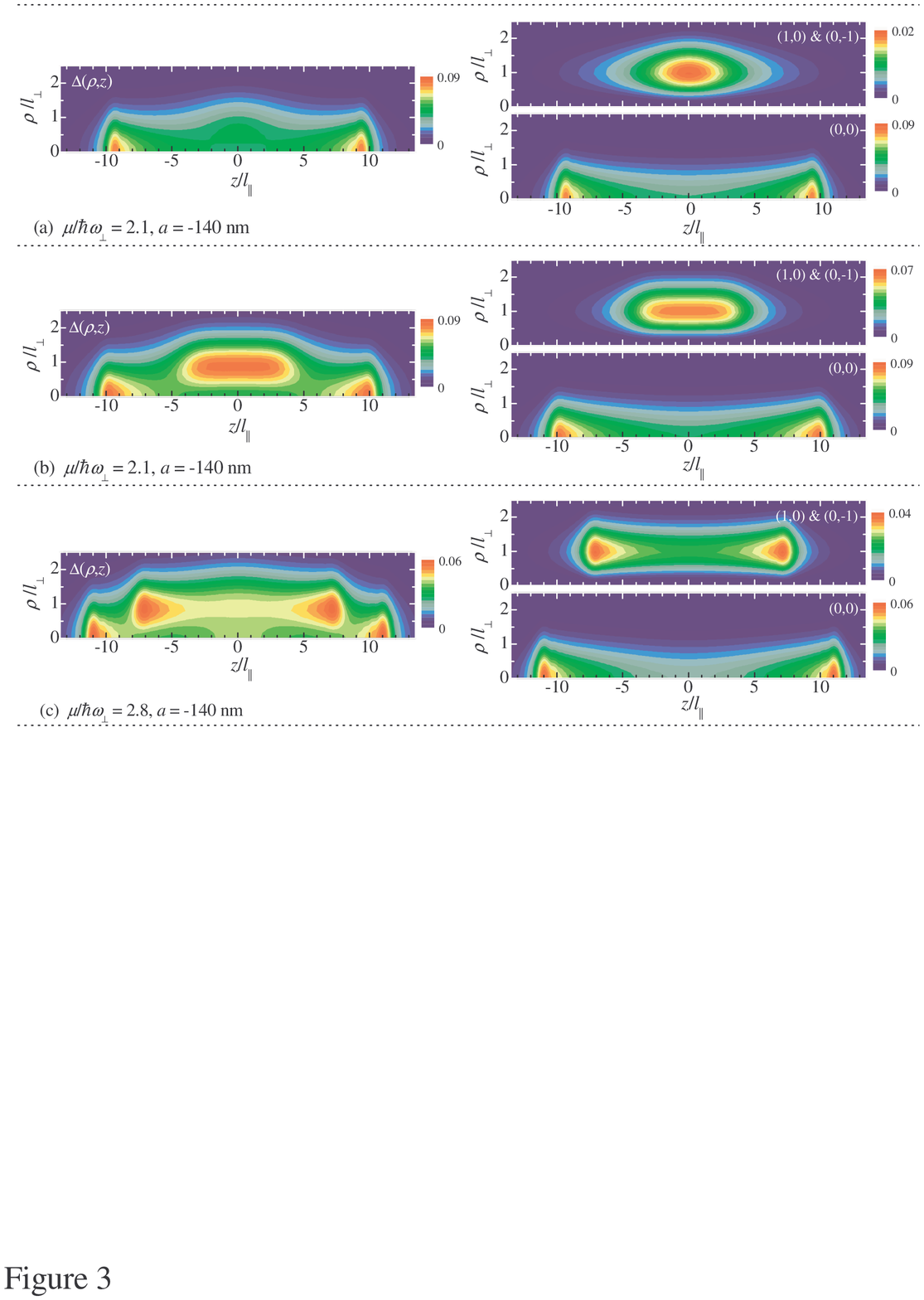}
\caption{(a) Contour plots of the order parameter $\Delta(\rho,z)$~[in
units of $\mu$, the left panel] together with the partial contributions
to the order parameter of the subbands with $(n,m)=(0,0)$ [the lower
right panel] and $(0,\pm1)$ [the upper right panel] for a cigar-shaped
superfluid Fermi gas at $\mu/\hbar\omega_{\perp} =1.9$ at $a=-140\,
{\rm nm}$; (b) and (c) show the same but for $\mu/\hbar\omega_{\perp}
=2.1$ and $2.8$.}
\label{fig3}
\end{figure}

In the previous paragraphs we considered the impact of the
multi-subband structure on the off-diagonal longitudinal behavior of
$\Psi({\bf r},{\bf r}')$. It is of interest to check what happens
with the diagonal function $\Psi({\bf r},{\bf r})$ that controls the
order parameter $\Delta({\bf r})=g\Psi({\bf r},{\bf r})$ and, so,
the spatial distribution of the fermionic condensate. In
Fig.~\ref{fig3} the contour plots of $\Delta(\rho,z)$~(calculated
for $T=0$ and given in units of $\mu$) are shown together with the
partial contributions of the relevant subbands at $a = -140\,{\rm
nm}$ for $\mu/\hbar\omega_{\perp}= 1.9,\,2.1$ and $2.8$. The left
contour plot of each panel shows the order parameter whereas the two
right contour plots depicts the partial contributions to
$\Delta({\bf r})$ of the subbands with $(n,m)=(0,0)$~[lower] and
$(n,m)=(0,\pm1)$~[upper]. At $\mu/\hbar \omega_{\perp}= 1.8$, see
panel (a), the quantum-size resonance associated with the two
degenerating subbands $(0,\pm1)$ begins to develop. So, the
longitudinal distribution of the order parameter is mainly
determined by the states with $(n,m)=(0,0)$. As a result, the
distribution of the condensate distribution is rather extended in
the $z$ direction, i.e., approximately from $-11\,l_{||}$ to $11\,
l_{||}$. Two points of enhancement of the order parameter can be
seen next to the left and right edges of the longitudinal condensate
distribution: at $z/l_{||} =-9.3$ and $9.3$, respectively. These
local enhancements are signatures of the BCS character of the
pairing in the quantum channel $(0,0)$: their $z$-coordinates are
solutions of the equation $M\omega_{||}z^2/2 =\hbar^2
(k^{(0,0)}_F)^2/2M$. Unlike $(n,m)=(0,0)$, subbands with $(0,\pm1)$
produce a contribution localized around $z=0$, which is in agreement
with the quasi-molecule character of the fermionic pairing in these
subbands. In panel (a) such a contribution does not have a
significant effect on the longitudinal distribution of $\Delta({\bf
r})$. However, when the bottoms of subbands $(0,\pm1)$ cross $\mu$,
the corresponding shape resonance develops and the condensate
distribution acquires a clear bimodal character, as seen from panel
(b). Here the main contribution to the pairing correlations comes
from the states with $(n,m)=(0,\pm1)$, which gives rise to a
significant enhancement of the order parameter around $z=0$, i.e.
from $z=-3\,l_{||}$ to $3\,l_{||}$. Such a localization along the
$z$ direction disappears when the perpendicular level
$2\hbar\omega_{\perp}$ goes significantly below $\mu$, as seen from
Fig.~\ref{fig3}(c). Here we arrive at the BCS picture of the
fermionic pairing in subbands $(0,\pm1)$, similar to that of
$(n,m)=(0,0)$.

\begin{figure}
\includegraphics[width=0.9\linewidth]{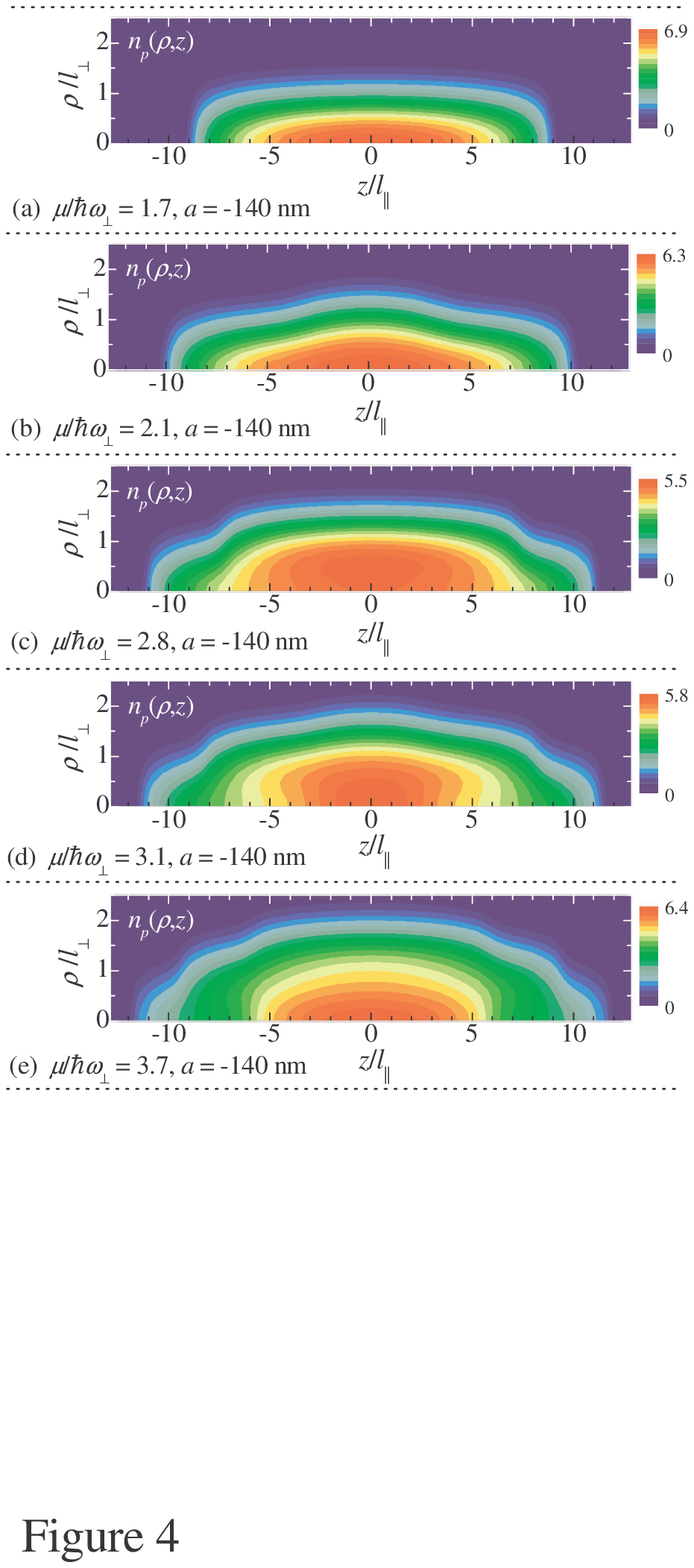}
\caption{Contour plots of the position dependent particle density
$n_p(\rho,z)$~(in units of $10^{12}\,{\rm cm}^{-3}$) for
$\mu/\hbar\omega_{\perp}=1.7$ (a), $\mu/\hbar\omega_{\perp}=2.1$
(b), $\mu/\hbar_{\perp}=2.8$ (c), $\mu/\hbar\omega_{\perp}=3.1$ (d)
and $\mu/\hbar\omega_{\perp}=3.7$ (e). For all the panels the
scattering length is $a=-140\,{\rm nm}$.} \label{fig4}
\end{figure}

Next we discuss important physical characteristics such as $T_c/T_F$
and $k_F|a|$ that are directly related to the problem of cooling the
Fermi gas\cite{bloch}. Typical values of $k_F|a|$ reached at the BCS
side of the Feshbach resonance are down to $\approx 1$ whereas the
corresponding values of $T_c/T_F$ fall into the interval $0.05$ -
$0.2$~(see Ref.~\onlinecite{bloch}). To check if our choice of $\mu$
is consistent with these experimental results, we need to estimate
$k_F$, the Fermi wavevector in the center of the trap. The simplest
way to estimate $k_F$ is to extract its value from the difference
$\mu-\hbar\omega_{\perp}$, i.e., to use $k_F
=\sqrt{\frac{2M}{\hbar^2} (\mu-\hbar\omega_{\perp})}$. Such an
estimate tells us that $k_F|a|$ varies from $0.56$ to $0.7$ at $a=
-140\,{\rm nm}$ when $\mu/\hbar\omega_{\perp}$ increases from $2$ to
$4$. At the same time, as follows from Fig.~\ref{fig1}(a),
$T_c/T_F$~($T_F=\hbar^2k^2_F/2M$) varies from $0.06$ down to
$0.03$~(see also Fig. 2 in Supplementary Information). For $a=-210
\,{\rm nm}$ we have $k_F|a|=0.84$ and $1.05$ at
$\mu/\hbar\omega_{\perp}=2$ and $4$, respectively. Here $T_c/T_F$
goes from $0.14$ down to $0.07$~(Supplementary Information, Fig. 2).
A more accurate estimate of $k_F$ can be found when calculating the
position dependent particle density $n_p(\rho,z)$~(see the relevant
formulas in Supplementary Information). Contour plots of
$n_p(\rho,z)$~[calculated at zero temperature for $a=-140\,{\rm nm}$
and given in units of $10^{12}\,{\rm cm}^{-3}$] are shown in
Fig.~\ref{fig4} for $\mu/\hbar\omega_{\perp}=1.7$~(a), $2.1$~(b),
$2.8$~(c), $3.1$~(d) and $3.7$~(e). In general, from Fig.~\ref{fig4}
one can extract particle densities close to those of the simplified
estimate. Indeed, based on $k_F=\sqrt{\frac{2 M}{\hbar^2}(\mu
-\hbar\omega_{\perp})}$, one finds that the mean particle density is
about $2\times 10^{12}\,{\rm cm}^{-3}$ at
$\mu/\hbar\omega_{\perp}=2$ and increases to $4\times 10^{12}\,{\rm
cm}^{-3}$ when $\hbar\omega_{\perp}$ approaches $4$. From
Fig.~\ref{fig4} it follows that the mean density in the center of
the trap is about $4$ - $5\times 10^{12}\,{\rm cm}^{-3}$. However,
it does not show any systematic increase with decreasing
$\omega_{\perp}$. As follows from Fig.~\ref{fig4}, $n_p(\rho,z)$ in
the center of the trap oscillates when changing $\omega_{\perp}$.
Such oscillations are due to an interplay of the relevant
single-particle subbands whose number changes by one each time when
$\mu/\hbar\omega_{\perp}$ approaches an integer. Now, taking
$n_p=4.5\times 10^{12}\,{\rm cm}^{-3}$ in the center of the trap and
using $a=-210\,{\rm nm}$, we obtain that $k_F|a|=1.07$ and $T_c/T_F$
varies between $0.07$ and $0.09$ (see Fig. 3, Supplementary
Information). Thus, both ways of estimating $k_F|a|$ and $T_c/T_F$
give experimentally attainable values. We remark that typical
densities of ultracold trapped alkali-metal gases are about
$10^{12}$ - $10^{15}\,{\rm cm}^{-3}$~(see Ref.~\onlinecite{bloch}).
We would also like to note that the quantum-size oscillations of the
coherence length are much more pronounced for smaller couplings,
i.e., for  $k_F|a| \ll 1$. However, for such couplings the ratio
$T_c/T_F$ becomes significantly below its experimentally attainable
lower bounds $0.05$ - $0.2$.

We also note that small quantum-size oscillations in $n_p(\rho,z)$
are both qualitatively and quantitatively different from those
observed in the fermionic condensate and, so, can in no way explain
the qualitative changes in the pairing correlations discussed above.
Although the common origin of oscillations in both single- and
many-body characteristics is the formation of the subband structure,
the superfluid quantum-size oscillations cannot be understood by
looking at the single-particle density $n_p(\rho,z)$. For example,
when the system goes through the first resonance, i.e.,  $1.7 <
\mu/\hbar \omega_{\perp} < 2.8$, the single-particle density in the
center of the trap changes by about $20\%$, see Fig.~\ref{fig4}.
Based on the homogeneous BCS theory, one can estimate that the
corresponding change in the BCS coherence length is only about
$7\%$, which is far smaller than observed in our calculations for
$a=-140\,{\rm nm}$.

Another interesting quantity to discuss is the product $k_F\xi_0$
that is used to to check the evolution of the fermionic condensate
at the BCS-BEC crossover. According to the paper\cite{Pistolesi94},
the intermediate region of a BCS-BEC crossover is approached when
$1/\pi < k_F\xi_0 < 2\pi$. This criterion corresponds, for a uniform
3D system, to the domain $-1 < (k_Fa)^{-1} < 1$, as discussed in
Ref.~\onlinecite{Perali02}. The upper boundary can be rewritten as
$\xi_0 =\lambda_F$. When $\xi_0 < \lambda_F$, a condensed fermionic
pair has an average size of the order of or less than the
interparticle spacing. Such a pair acquires a local character: it
weakly, or even does not, overlap with other fermionic pairs. The
lower boundary $k_F\xi_0 = 1/\pi$ sets the BEC regime (i.e., for
$k_F\xi_0 < 1/\pi$) with the formation of stable point-like
molecules. Let us now consider how the product $k_F\xi_0$ changes
with $\mu/\hbar\omega_{\perp}$ in a quasi-1D fermionic condensate.
The most interesting is the case of $a=-140\,{\rm nm}$. Here
$k_F\xi_0$ is well above $2\pi$ except for the points of the
size-dependent drops of $\xi_0$, where this product reaches values
between $4$ and $6$~(see Supplementary Information). It means that
each time when the bottom of a single-particle subband crosses
$\mu$, we approach the intermediate region of our BCS-BEC-like
crossover driven by quantum-size effects. As for the
subband-dependent length $\xi^{(nm)}_0$, it is always well below
$2\pi$ in the vicinity of the corresponding quantum-size resonance.
For instance, $k_F\xi^{(0,\pm1)}_0$~(taken at
$\mu/\hbar\omega_{\perp}=2$) is about $2.0$ - $2.5$~(see
Supplementary Information). This is of importance because, e.g., for
$\mu/\hbar \omega_{\perp}=2$ about $70$ - $75\%$ of the condensed
pairs come from the subbands with $(n,m)=(0,\pm1)$, as already
mentioned above. Note that for the subband-dependent coherence
length it looks more natural to check the product $k^{(n,m)}_F
\xi^{(nm)}_0$ that gives significantly smaller values as compared to
$k_F\xi^{(nm)}_0$. However, in the vicinity of a quantum-size
resonance, $k^{(nm)}_F$ associated with the resonant subband
significantly underestimates the mean longitudinal single-particle
momentum due to the neglect of the contribution from pairing
correlations. Moreover, $k^{(nm)}_F$ is not defined when the bottom
of a subband is positioned above $\mu$. So, a proper generalization
of the criterion $1/\pi < k_F\xi_0 < 2\pi$ for the subband-dependent
BCS-BEC crossover can be of interest.

Several remarks are in order about the ultraviolet regularization
used in our calculations. As mentioned above, to arrive at an
ultraviolet finite approach, we adopt the regularization by
subtracting $1/(2T_{nmj})$ in Eqs.~(\ref{3D})~(taken with ${\bf k}
\to \nu = \{n,m,j\}$), (\ref{Psi}) and (\ref{psiz}), which is a
simple and straightforward extension of the regularization for
homogeneous superfluid systems to our spatially nonuniform case. In
fact, our procedure is a simplified version of a comprehensive
scheme for spatially nonuniform systems suggested and investigated
in Ref.~\onlinecite{bruun}. Such a simplification is very similar to
the ultraviolet regularization used in Ref.~\onlinecite{strinati}.
We stress that our basic conclusions are not sensitive to the
specific form of the ultraviolet regularization. In particular,
almost the same results are obtained with a simple cut-off
procedure, i.e., the term $-1/(2T_{nmj})$ is removed from
Eqs.~(\ref{3D}), (\ref{Psi}) and (\ref{psiz}) and the summation in
the relevant expressions runs over the single-particle states with
$|T_{nmj}-\mu| < \mu$, see Fig. 4 in Supplementary Information. A
similar ultraviolet cut-off is often used in papers dealing with
superfluidity in nuclei and uniform atomic condensates, see, e.g.,
refs.~\onlinecite{bruun1,heis,torma}.

Concluding this section, we note that the longitudinal shrinking of
condensed fermionic pairs driven by quantum-size effects in a
cigar-shaped fermionic condensate can be investigated experimentally
via rf-spectroscopy\cite{ket}. Another possibility is to exploit the
presence of a coherent superposition of the subband-dependent
fermionic condensates. When a shape resonance develops, one of them
has a molecule-like character and the others are of the BCS type.
Such fermionic condensates will evolve in a different manner when
switching off the longitudinal confinement. Supplemented by
measurements of the quantum-size oscillations of the basic
quantities, i.e., $T_c$ and relevant pairing gaps (the gaps can be
probed by the rf-spectroscopy), such experiments could give useful
information about the quantum-size driven reconstruction of the
fermionic pairing in a confined condensate.

\section{Quantum superconducting nanowires}

The effect in question is universal and can also be expected for
high-quality superconducting nanowires, when disorder is small
enough and, so, scattering on imperfections do not shadow the
formation of single-electron subbands due to the transverse
quantization of electron motion. For conventional superconducting
materials, e.g., ${\rm Al}$ or ${\rm Sn}$, the bulk
(zero-temperature) excitation gap $\Delta_{\rm bulk}$ varies
from\cite{degen} $\sim 0.1$ to $\sim 1.0\,{\rm meV}$ and, so, the
inter-subband energy spacing $\delta \approx \frac{\hbar^2}{2m_e}
\frac{\pi^2}{d^2}$~(with $d$ the nanowire diameter) is of the same
order or larger when $d<20-40\,{\rm nm}$. For these diameters
quantum-size effects on the superconducting properties become of
importance. In particular, recently found width-dependent systematic
shift of $T_c$ to upper values in superconducting aluminum/tin
nanowires\cite{tian,zgi,alt} have been attributed to these
quantum-size effects\cite{shan}. We remark that at present the
narrowest fabricated superconducting nanowires are aluminum samples
with width down to $8$ - $11\,{\rm nm}$~(see Ref.~\onlinecite{alt}).

All the formulas for the superconducting condensate in a quantum
wire are similar to those of the previous section. However the
relevant single-electron wave functions will be different. The
electron motion is not quantized in the $z$ direction (we deal with
periodic boundary conditions with unit cell of length $L$) and, so,
for $\psi_i(z,z')=\psi_i(z-z')$ we have (at $T=0$)
\begin{equation}
\psi_i(z-z')=\frac{1}{L}\sum\limits_k\,\frac{\Delta_i}{2E_{ik}}
e^{\imath k (z-z')} = \frac{1}{\pi} \int\limits_{k_1}^{\,k_2}
\!\!{\rm d}k\,\frac{\Delta_i}{2E_{ik}}\,e^{\imath k (z-z')},
\label{wire}
\end{equation}
where $i$ stands for the set of quantum numbers controlling the
motion perpendicular to the nanowire; $\Delta_i$~(chosen real) is
the subband pairing gap that does not depend on the longitudinal
wavevector $k$~(due to periodic boundary conditions in the
$z$-direction); and $E_i=\sqrt{\varepsilon^2_{ik} + \Delta^2_i}$,
where $\varepsilon_{ik}$ is the single-electron energy measured from
the chemical potential, i.e., $\varepsilon_{ik}=\varepsilon_{i}
+\frac{\hbar^2 k^2}{2m_e} -\mu$ with $\varepsilon_i$ the discrete
perpendicular single-electron level. As seen from Eq.~(\ref{wire}),
there is no regularization term similar to that of Eq.~(\ref{psiz}).
But instead, the sum in Eq.~(\ref{wire}) runs only over the
single-particle states in the Debye window, i.e., with
$|\varepsilon_{ik}| <\hbar\omega_D$, where $\omega_D$ is the Debye
frequency appearing in the problem due to the phonon mediated
attraction between electrons. This is why the lower $k_1$ and upper
$k_2$ cut-off momenta appear in the integral in Eq.~(\ref{wire}).
For $k=k_1$, $\varepsilon_{ik}$ crosses the lower boundary of the
Debye window, i.e., $k_1$ is a nonnegative solution of
$\varepsilon_{ik} = -\hbar\omega_D$. This is when the subband bottom
$\varepsilon_i$ is situated below $\mu-\hbar\omega_D$. When the
subband bottom goes above $\mu-\hbar\omega_D$, $k_1$ is set to zero.
In turn, $k_2$ is a nonnegative solution of $\varepsilon_{ik}=\hbar
\omega_D$ provided that the subband bottom is below the upper
boundary of the Debye window. When $\varepsilon_i > \mu +
\hbar\omega_D$, we get $k_{\rm up}= 0$ and, so, such a subband does
not contribute to the superconducting characteristics. A major
contribution to the integral in Eq.~(\ref{wire}) comes from the
states in the vicinity of the minimum of $|\varepsilon_{ik}|$ so
that in most interesting cases one can simply put $k_1=0$ and $k_2
\to +\infty$~(the resulting integral is perfectly convergent). It
means that the presence of the cut-off does not significantly
influence the longitudinal properties of a quasi-1D superconducting
condensate and, in turn, our conclusions. Based on Eq.~(\ref{wire}),
one can calculate $\Psi({\bf r},{\bf r}')$ from Eq.~(\ref{Psi}),
where $n,m$ are replaced by $i$ and $\vartheta_i$ is no longer the
eigenfunction of the 2D isotropic harmonic oscillator but the
single-electron wave function corresponding to the transverse
quantum number $i$.

When inserting $k_1=0$ and $k_2= +\infty$ into the integral in
Eq.~(\ref{wire}), the decay length of $\psi_{nm}(z-z')$ can be
calculated analytically by using the contour integration in the
complex plane (this length is proportional to the longitudinal
subband-dependent coherence length $\xi^{(i)}_0$). The integrand in
Eq.~(\ref{wire}) has four singular points (the square-root branch
points) with the same absolute value for the imaginary part. For,
say, positive $z-z'$, the contour should be closed in the upper half
plane and distorted to encircle the cut between the two upper
singular points having the same imaginary part $[m_e(\sqrt{\mu^2_i +
\Delta_i^2} -\mu_i)]^{1/2}/\hbar$, where $\mu_i=\mu -\varepsilon_i$
is the chemical potential measured from the subband bottom. This
imaginary part controls the decay of $\psi_i(z-z')$ with increasing
$z-z'$, so that we obtain
\begin{equation}
\xi_0^{(i)} \propto \frac{\hbar}{\sqrt{m_e}}\left[\sqrt{\mu^2_i +
\Delta_i^2}-\mu_i\right]^{-1/2}. \label{subleng}
\end{equation}
When $\mu_i/\Delta_i \gg 1$, i.e., the ratio of the longitudinal
kinetic energy to the interaction energy in the corresponding
subband is large enough, we arrive at the conventional result for
the BCS coherence length, i.e., $\xi_0^{(i)} \propto \hbar
v_i/\Delta_i$, with $v_i = \sqrt{2 \mu_i/m_e}$ the longitudinal
subband-dependent Fermi velocity. This is for a subband with the
bottom far below the Fermi surface so that $v_i \approx v_F$, with
$v_F=\sqrt{2\mu/m_e}$. As $\Delta_i \propto T_c$, we get
$\xi^{(i)}_0 \propto \hbar v_F/T_c$, which is in agreement with our
results for $\xi^{(0,0)}_0$ of a cigar-shaped fermionic condensate
in Fig.~\ref{fig1}. At the point of a quantum-size superconducting
resonance, when $\mu_i \rightarrow 0$, Eq.~(\ref{subleng}) reduces
to a completely different expression, i.e., $\xi_0^{(i)} \approx
\hbar/(m_e \Delta_i)^{1/2}$. This is also fully consistent with our
numerical results given in Fig.~\ref{fig1} (see the discussion in
the previous section about $xi_0^{nm}$ for $(n,m)=(1,0),\,(2,0)$ and
$(3, 0)$ at points $\mu/\hbar \omega_{\perp}=2,\,3$ and $4$,
respectively). When $\mu_i < 0$ and $|\mu_i| \gg \Delta_i$, one
obtains $\xi_0^{(i)} \propto \hbar/(2m_e|\mu_i|)^{1/2}$. Here the
dependence of $\xi^{(i)}_0$ on the subband pairing gap
$\Delta_i$~(and, so, on $T_c$) disappears entirely. The same was
found in the previous section for $\xi^{(3,0)}_0$ for
$\mu/\hbar\omega_{\perp} < 3.5$, see Fig.~\ref{fig1}.

\begin{figure}
\includegraphics[width=0.82\linewidth]{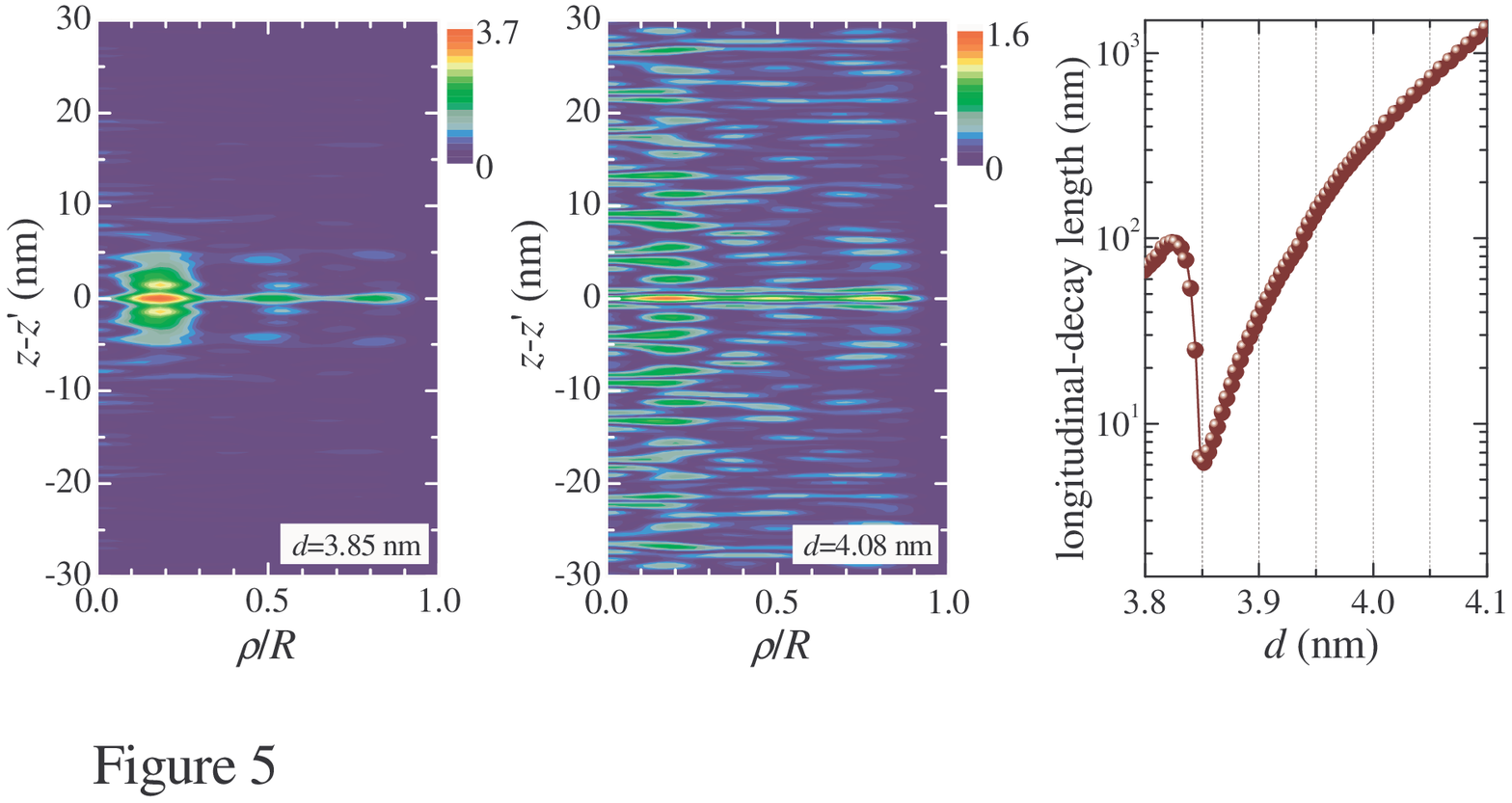}
\caption{Superconducting nanocylinder: (a) contour plot of
$g\Psi(\rho,\varphi,z;\rho,\varphi,z')$~(given in units of the bulk
gap $\Delta_B=0.25\,{\rm meV}$) in the presence of the
superconducting resonance at $d=3.85\,{\rm nm}$; (b) the same as in
the previous panel but now for the non-resonant diameter
$d=4.08\,{\rm nm}$; (c) the longitudinal decay length for the
distribution $\Psi(\rho,\varphi, z;\rho,\varphi, z')$~(it is
proportional to $\xi_0$) versus the diameter near the resonance
point $d=3.85\,{\rm nm}$.} \label{fig5}
\end{figure}

The simple analytical structure of Eq.~(\ref{subleng}) highlights a
strong similarity of our results with the BCS-BEC crossover driven
by a change of the strength of the pair interaction: $\mu_i$ and
$\Delta_i$ in Eq.~(\ref{subleng}) can simply be replaced by the
chemical potential $\mu$ and the gap $\Delta$ to reproduce the
evolution of the size of the condensed fermionic pairs through the
region of the crossover driven by the Feshbach resonance (see, e.g.,
Ref.~\onlinecite{bloch}). However, there is also an important
difference: in our case a trend similar to a BCS-BEC crossover is
found for the resonant single-particle subband whose bottom is close
to the Fermi surface. Other relevant subbands remain in the ordinary
BCS regime. Thus, as already mentioned above, at the point of a
quantum-size resonance the pair condensate is governed by a coherent
superposition of the quasi-molecule quantum channel with a set of
ordinary BCS channels. Resonant channels significantly contribute to
the condensation energy. For example, this is about $70\%$ and
$50\%$ for a superfluid cigar-shaped Fermi gas at
$\mu/\hbar\omega_{\perp}=2$ and $3$, respectively (see the previous
section). For superconducting cylindrical nanowires with diameters
less than $8-10\,{\rm nm}$ the contribution of the resonant subbands
at a quantum-size superconducting resonance is typically about
$60-70\%$. This makes it possible to conclude that sufficiently
narrow superconducting nanowires at the resonant points are mainly
governed by the quasi-molecule channel of the fermionic condensate.
As an illustration, Figs.~\ref{fig5} (a) and (b) show contour plots
of the off-diagonal superconducting order parameter $g \Psi(\rho,
\varphi,z; \rho,\varphi,z')$~(the transverse coordinates of two
electrons are taken the same) for a cylindrical superconducting
nanowire at diameters $d=3.85\,{\rm nm}$~[panel (a) is for the case
of a quantum-size superconducting resonance] and $d=4.08\,{\rm
nm}$~[panel (b) is for the non-resonant case]. Such a representation
is very convenient because it provides us with information of not
only the longitudinal correlations but, in addition, on the
(diagonal) order parameter $\Delta(\rho) =g\Psi({\bf r},{\bf r})$.
Unlike the previous case of the cigar-shaped trap, the order
parameter does not depend on $z$ due to periodic boundary conditions
in the $z$ direction. As seen from Figs.~\ref{fig5}(a) and (b), the
longitudinal distribution of electrons in a condensed pair
significantly shrinks at $d=3.85\,{\rm nm}$ as compared to
$d=4.08\,{\rm nm}$. This is because at $d=3.85\,{\rm nm}$ the
bottoms of the two single-electron subbands with the quantum numbers
$i=(n,m)=(5,\pm1)$ and $(1,\pm11)$ are situated in the vicinity of
the Fermi surface~(due the cylindrical shape the relevant transverse
quantum numbers are again $n$ and $m$, the radial and azimuthal
numbers, respectively). The corresponding quantum-size
superconducting resonance is responsible for a significant
enhancement of the order parameter $\Delta(\rho)$ as compared to its
bulk value (for our parameters it is $\Delta_B = 0.25\,{\rm nm}$,
see Supplementary Information) and for the main contribution of
these subbands to its averaged value (about $70\,\%$). When
increasing the diameter, these subbands shift down in energy, the
corresponding resonance decays and, so, one can see a rather
extended distribution along the $z$ direction at $d=4.08\,{\rm nm}$,
see panel (b). Fig.~(\ref{fig5})(c) demonstrates the
longitudinal-decay length of $\Delta(\rho,z-z')$ calculated from our
results by a numerical exponential fit for $\rho/R=0.18$. As seen,
the variation of the length (it is proportional to the longitudinal
pair size $\xi_0$) for diameters from $d=3.85\,{\rm nm}$ to
$4.08\,{\rm nm}$ is two-three orders of magnitude. It is
significantly more pronounced than the drops in $\xi_0$ found in the
previous section for an ultracold quasi-1D Fermi gas. The reason is
as follows. From our analytical results it follows that the
resonance-driven decrease of the longitudinal coherence length is
governed by a factor $\approx \hbar k_F/\sqrt{m\Delta}$, where
$\Delta$ is the typical energy gap; $m$ is the relevant particle
mass (e.g., $m=m_e$ for electrons and $m = M$ for lithium atoms);
and $k_F$ can be estimated as $\sqrt{\frac{2M}{\hbar^2} (\mu-\hbar
\omega_{\perp})}$ for a cigar-shaped Fermi gas whereas for a
metallic nanowire we have $k_F=\sqrt{2m_e/\hbar^2}$. Then, one can
find that the reason for the significantly more pronounced drop of
the longitudinal BCS coherence length in the superconducting
nanowire is the value of $k_F$ which is more than three orders of
magnitude larger in metals. The effect of an increase of the typical
excitation gap in metals by seven orders of magnitude as compared to
a trapped atomic condensate is strongly weakened by the $10^4$-drop
of the particle mass and, in addition, by the presence of the
product $m\Delta$ under the square root. For instance, one can
simply compare $\xi^{(i)}_0 = \hbar v_i/\Delta_i$ with $\xi^{(i)}_0
=\hbar/(m_e \Delta_i)^{1/2}$ for typical metallic parameters of
weak-coupling superconductors: $v_i \sim 1 \times 10^6\,-\,2 \times
10^6\,{\rm m/s}$ and $\Delta_i \approx 0.2\,{\rm meV} $~(see.e.g.,
Ref.~\onlinecite{degen}). This allows one to find that $\xi^{(i)}_0$
drops at a resonant point by two-three orders of magnitude, from the
micron to the nanometer scale, i.e., down to  $\xi^{(i)}_0 \sim
20\,{\rm nm}$. Then, taking account of the fact that the diagonal
order parameter $\Delta(\rho) = \Delta(\rho, z-z')|_{z-z'=0}$ is
$3-4$ times enhanced as compared to bulk at $d=3.85\,{\rm nm}$, one
can eventually obtain a drop shown in Fig.~\ref{fig5} (c). We note
that when moving below diameter $3.85\,{\rm nm}$, the longitudinal
decay-length first increases with decreasing $d$ but then, when
reaching $d=3.82\,{\rm nm}$, it begins to drop. This is the effect
of another quantum-size resonance situated at $d = 3.72\,{\rm
nm}$~(see Ref.~\onlinecite{shan1}). For diameters larger than $3.85
\,{\rm nm}$, the next quantum-size resonance develops at
$d=4.2\,{\rm nm}$.

We would like to remark that the above results for superconducting
nanowires were calculated in the clean limit. In the presence of
surface imperfections and disorder of real metallic nanowires the
quantum-size oscillations of the longitudinal coherence length will
be smoothed. However, such giant drops of $\xi_0$ can hardly be
completely washed out (see discussion in Supplementary Information).
Yet, it will be difficult to observe details of oscillations of the
BCS coherence length due to fluctuations of the confinement
dimensions of real samples. To capture the details, it is more
promising to use materials with lower charge-carrier densities
(e.g., semimetals and doped semiconductors). Such materials have
larger $\lambda_F$ that controls the number of the occupied
transverse modes and the width of the superconducting resonances
(this width can be defined as a maximal variation of the nanowire
diameter with no significant effect on a resonant enhancement). We
can expect that for $\lambda_F \gg 1\,{\rm nm}$ smoothing of
quantum-size oscillations due to width fluctuations will be less
significant.

\section{Conclusion}

Since the classical paper by Cooper\cite{cooper} it is well-known
that the configuration of the phase space available for scattering
of time-reversed fermions plays a crucial role for the formation of
condensed fermionic pairs. Indeed, only a strong enough attractive
interaction between fermions with opposite spin in 3D is able to
produce a two-body bound state in vacuum. However, when scattering
of fermions is influenced by the exclusion of a filled Fermi sea,
i.e., the available phase space is restricted by exclusion of the
single-particle states inside the Fermi sea, we arrive at the Cooper
instability resulting in the formation of weakly bound in-medium
pairs of fermions for arbitrary strength of the attractive
interaction. Restricting the phase space by removing the filled
Fermi sea, one actually removes long range contributions to the
Cooper-pair wave function, which, say, ``encourage" fermions to form
in-medium bound states. Our present results show that the additional
reconfiguration of the phase space due to quantum confinement, such
that the band of single-particle states splits up into a series of
subbands, can further modify the scenario of pairing. The formation
of multiple subbands due to quantum confinement results in a
significant redistribution of the kinetic energy between confined
and unconfined directions. In particular, for a quasi-1D fermionic
condensate the subband-dependent ratio of the longitudinal kinetic
energy to the interaction energy drops down to almost zero when the
bottom of a subband crosses the Fermi level (this is rather similar
to superconducting semimetals with the chemical potential driven by
superconducting correlations below the bottom of the conduction
band\cite{eagles}). The longitudinal Fermi motion of fermions in
such a subband is depleted, which results in a drop of the
longitudinal size of condensed fermionic pairs. So, each time when
the bottom of a single-particle subband passes through the Fermi
level, a quantum-size resonance develops and the
superfluid/superconducting system exhibits trends similar to the
well-known BCS-BEC crossover driven through the Feshbach resonance
in ultracold Fermi gases in 3D traps (see, e.g.,
Ref.~\onlinecite{bloch}). In this case condensed fermionic pairs
behave, to a great extent, similar to condensed boson-molecules. For
instance, this is reflected in a clear bimodal spatial distribution
of a harmonically trapped quasi-1D fermionic condensate. A
contribution governed by the quasi-molecule channel associated with
the subband whose bottom is situated in the vicinity of the Fermi
surface is strongly localized in the center of a trap. Other quantum
channels are due to single-particle subbands with bottoms far below
the Fermi surface. They are in the BCS regime and yield an extended
longitudinal distribution of the fermionic condensate typical of
Cooper pairs. When the number of relevant single-particle subbands
increases, the role of quantum-size effects is diminished and,
finally, all related phenomena disappear. One can expect that the
impact of the subband formation on superfluid/superconducting
properties is of importance when the inter-subband energy spacing is
larger than or close to $\approx k_B T_c$. We remark that similar
physics can be expected for quasi-2D fermionic condensates.

\begin{flushleft}
{\bf \small Acknowledgements} This work was supported by the
Flemish Science Foundation (FWO-Vl) and the Belgian Science
Policy (IAP). M.D.C. acknowledges support of the European
Community under a Marie Curie IEF Action (Grant Agreement No.  PIEF-GA-2009-235486-ScQSR).
\end{flushleft}

\appendix

\begin{flushleft}
{\Large \bf Supplementary information}\\
\end{flushleft}

\section{Quasi-$1D$ superfluid Fermi gas}
\label{sec_sup1}

\begin{figure}[b]
\resizebox{0.8\columnwidth}{!}{\rotatebox{0}{
\includegraphics{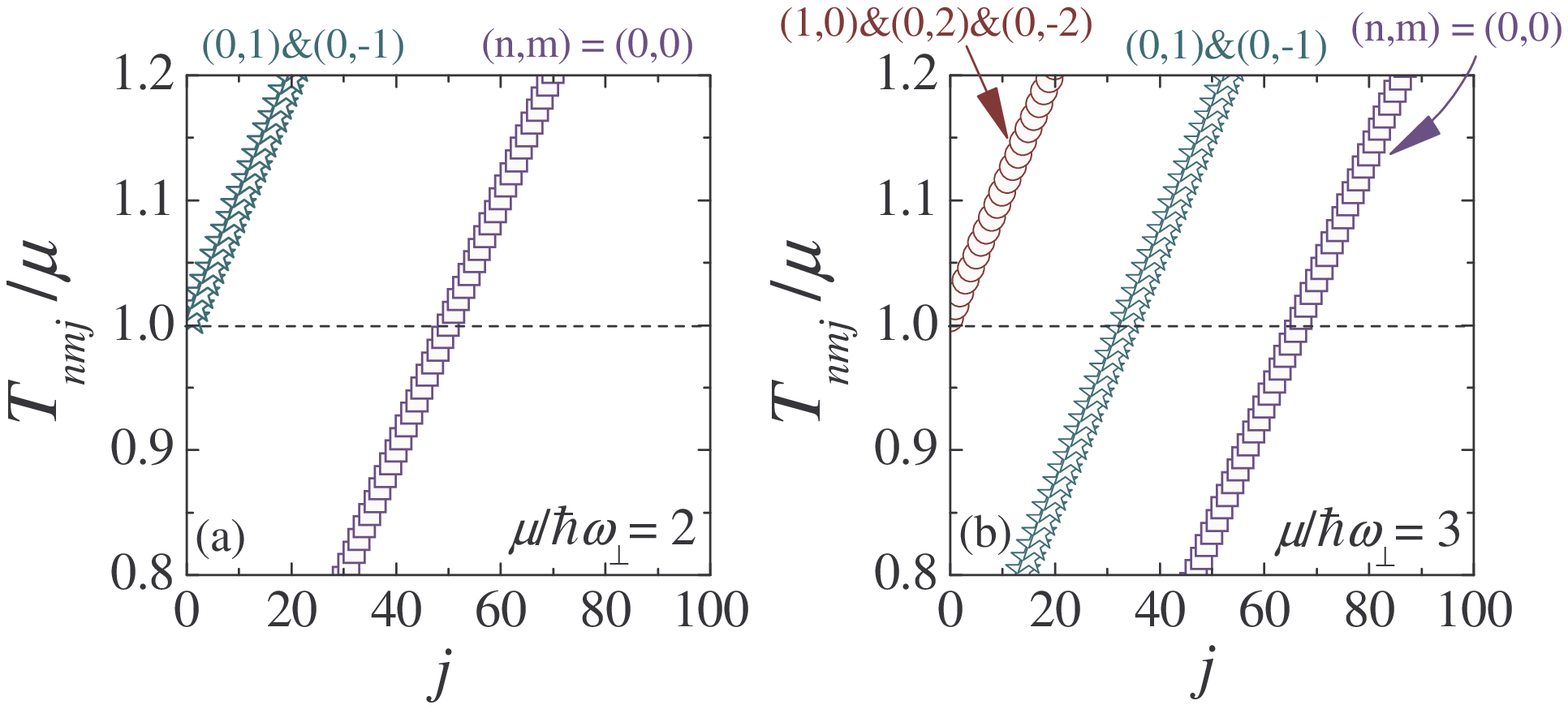}}}
\caption{The single-particle energy $T_{nmj}$ versus the longitudinal quantum number $j$ for different subbands $(n,m)$ at $\mu/\hbar\omega_{\perp}=2$~(a)
and $\mu/\hbar\omega_{\perp}=3$.} \label{fig6}
\end{figure}

Our investigation is focused on the quasi-1D Fermi gas, where the single-particle energies $T_{nmj}$~(with $j$ corresponding to the longitudinal motion) with the same $n$ and $m$ form a quasi-continuum. So, it is convenient to introduce single-particle subbands $(n,m)$ as seen from Fig.~\ref{fig6}. In panel (a) we
show $T_{nmj}$ versus the quantum number $j$ for $\mu/\hbar\omega_{\perp}=2$. In this case the bottom of subband $(0,0)$ is located far below the chemical potential $\mu$ whereas the bottom of two degenerate subbands $(0,\pm1)$ almost touches $\mu$~(it is higher only by $\hbar\omega_{||}/2= 0.005 \mu$, where we take $\mu=h \cdot 24\,{\rm kHz}$, see the discussion about the chosen parameters in the article). When decreasing $\omega_{\perp}$~(and keeping $\mu$ the same), more and more new subbands crosses the chemical potential. In particular, Fig.~\ref{fig6}(b) illustrates the case of $\mu/\hbar\omega_{\perp}=3$. Here the bottoms of subbands
$(0,0)$ and $(0,\pm1)$ are below $\mu$. The bottoms of the three degenerate subbands with $(n,m)=(0,\pm2)$ and $(1,0)$ almost touch $\mu$, which results in the appearance of a new quantum-size resonance.

When discussing in the article important physical parameters of the superfluid fermionic condensate such as $T_c/T_F$, $k_F|a|$, and $k_F\xi_0$, we invoked two different ways to estimate $k_F$ in the center of the trap. The simple one involves the chemical potential $\mu$ measured from the lowest single-particle level $\hbar\omega_{\perp} + \frac{1}{2} \hbar\omega_{||} \approx \hbar\omega_{\perp}$ and is based on the relation $k_F=\sqrt{\frac{2M}{\hbar^2} (\mu - \hbar \omega_{\perp})}$. This estimation is good enough when a trap is
almost isotropic, i.e., the trapping frequencies are close to each other in all directions. For the quasi-1D case, when $\omega_{\perp}\gg \omega_{||}$, the reliability of the above estimate can suffer from the anisotropy in the distribution of the kinetic energy in single-particle subbands. For instance, when
the subband bottom is situated in the vicinity of $\mu$, the longitudinal kinetic energy is suppressed as compared to the kinetic energy of the motion perpendicular to the system. While the subband bottom goes far below $\mu$, the above anisotropy disappears in favor of another: now the longitudinal motion prevails. So, the
resulting values of $k_F$ and $T_F = \hbar^2k^2_F/2m_e$ will be dependent on whether $k_F$ is estimated from the longitudinal kinetic energy or it is extracted from the energy of the perpendicular motion. Another issue of uncertainties is the choice of the relevant subband to extract $k_F$. Notice that the above simple
estimate is based on the longitudinal kinetic energy of the lowest single-electron subband.

\begin{figure}[tbp]
\resizebox{0.7\columnwidth}{!}{\rotatebox{0}{
\includegraphics{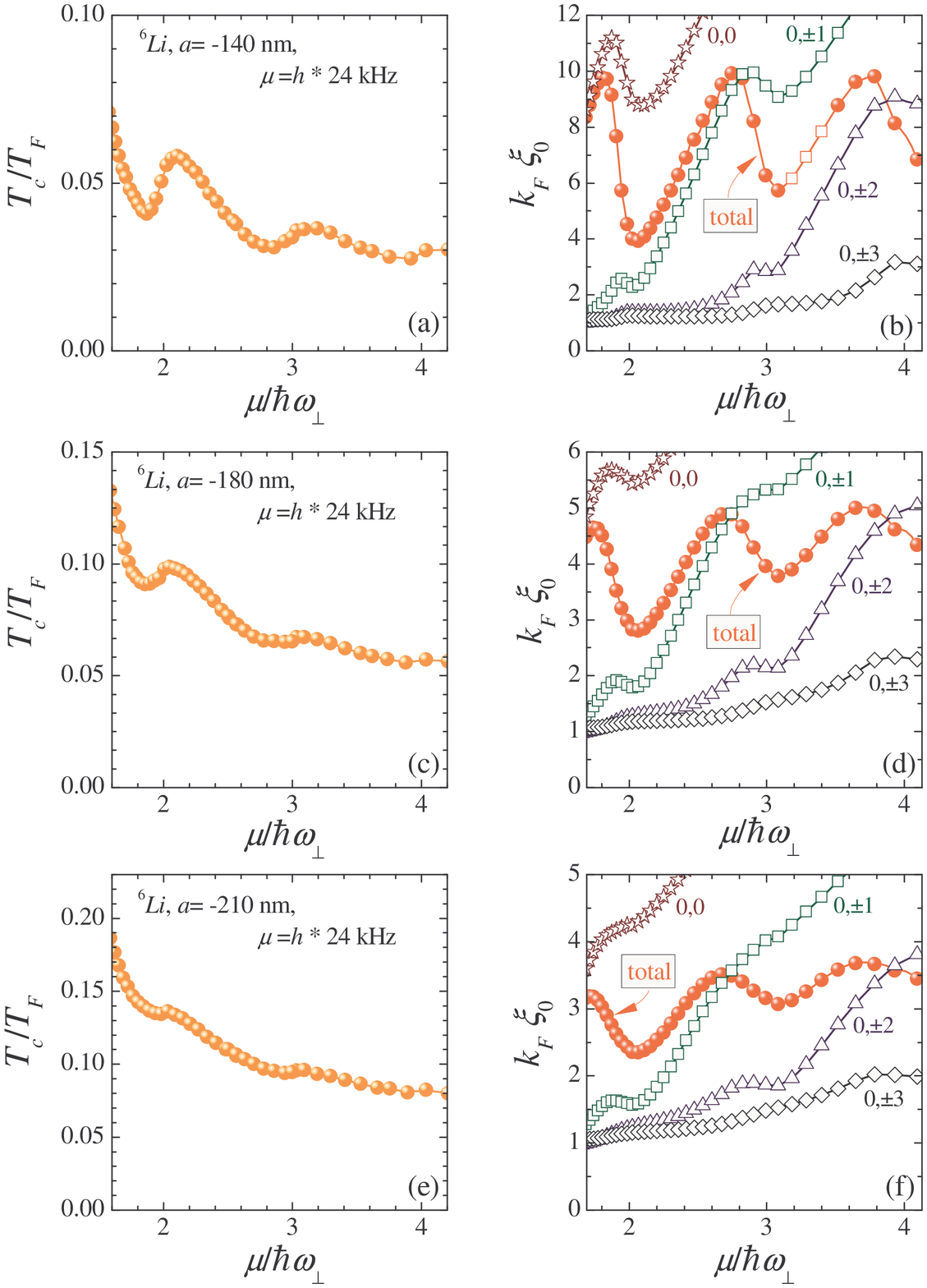}}}
\caption{$T_c/T_F$ and $k_F\xi_0$ as calculated from the results
given in Fig. 1 of the article on the basis of the simple estimate
of $k_F=\sqrt{\frac{2M}{ \hbar^2}(\mu-\hbar\omega_{\perp})}$~(with
$T_F =\hbar^2 k^2_F/2M$). Panels (a), (b) represent $a=-140\,{\rm
nm}$; (c) and (d) are for $a=-180\,{\rm nm}$; (e), (f) show the data
for $a=-210\,{\rm nm}$. The quantities $k_F\xi^{(nm)}_0$ for
$(n,m)=(0,0),\,(0,\pm1)\,(0, \pm2)$ and $(0,\pm3)$ are also plotted
to compare with $k_F\xi_0$.} \label{fig7}
\end{figure}

\begin{figure}[tbp]
\resizebox{0.7\columnwidth}{!}{\rotatebox{0}{
\includegraphics{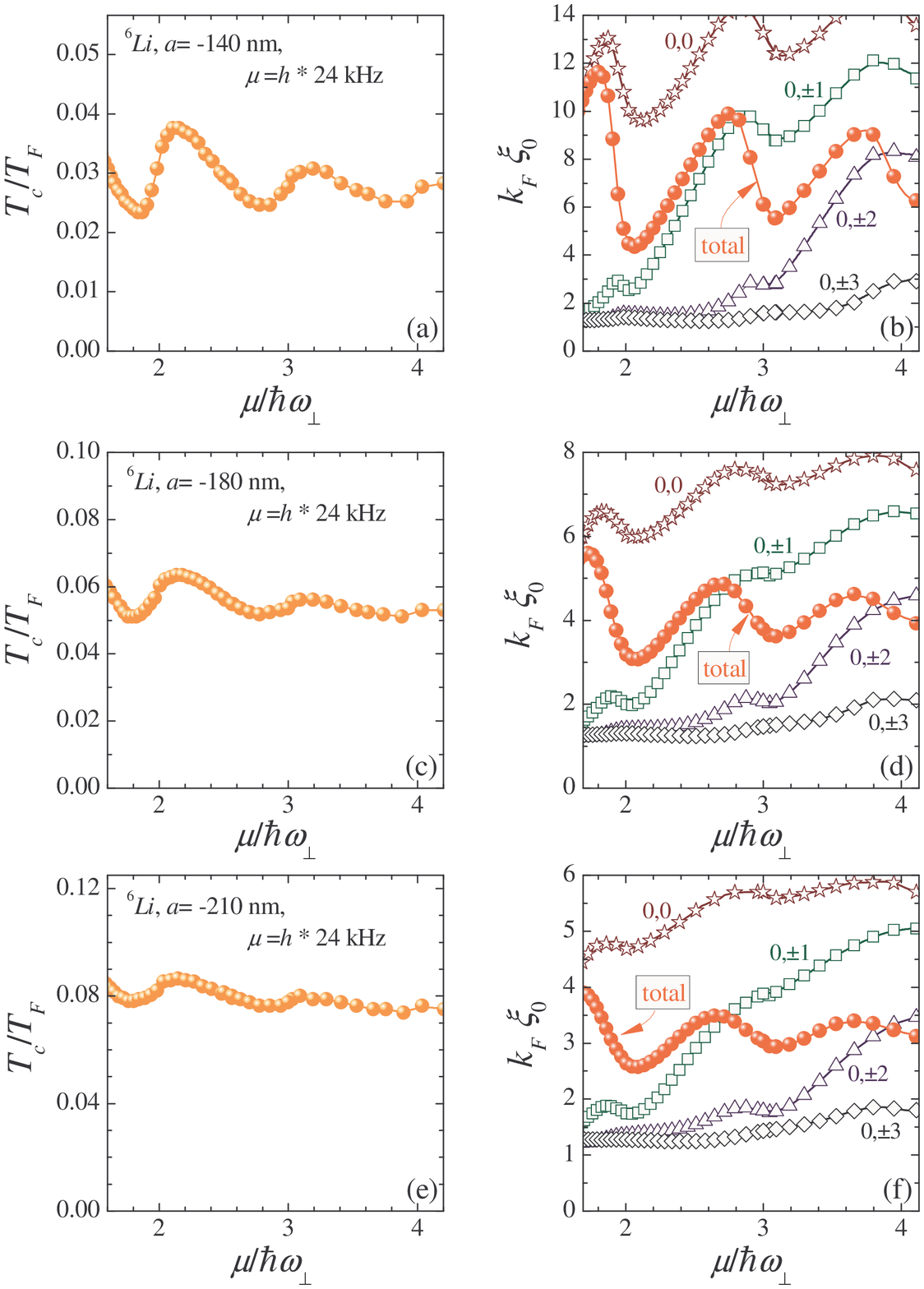}}}
\caption{The same as in the previous figure but now for $k_F$
extracted from the mean particle density in the center of the trap,
i.e., $n_p = 4.5 \times 10^{12}\, {\rm cm}^{-3}$ ($k_F = (3 \pi^2
n_p)^{1/3}$).} \label{fig8}
\end{figure}

A more accurate (but more time consuming) procedure to estimate the relevant $k_F$ in the center of the trap is to calculate the spatial particle distribution $n_e({\bf r})$ and, then, to find the mean particle density in the center of the trap. In the mean-field approximation the quantity $n_e({\bf r})$ is calculated through the coherence factors $u_{\nu}({\bf r})$ and $v_{\nu}({\bf r})$, see
Eq.~(3) in the article, i.e.,
\begin{equation}
n_p({\bf r}) = 2\sum\limits_{\nu} \Bigl[|v_{\nu}({\bf r})|^2\,(1-
f_{\nu}) + |u_{\nu}({\bf r})|^2\,f_{\nu}\Bigr],
\label{1}
\end{equation}
with $f_{\nu}=1/(e^{\beta E_{\nu}} + 1)$, the Fermi distribution of the bogolons. We note that for a trap with axial symmetry the density of particles depends only on the radial coordinate $\rho$ and the longitudinal coordinate $z$. Based on Eq.~(3) of the article, one can rewrite Eq.~(\ref{1}) as (at $T =0$)
\begin{equation}
n_p(\rho,z)=\sum\limits_{nmj}\Bigl(1-\frac{T_{nmj}-\mu}{E_{nmj}}
\Bigr)\,|\vartheta_{nm}(\rho,\varphi)|^2\,\chi^2_j(z),
\label{2}
\end{equation}
where $|\vartheta_{nm}(\rho,\varphi)|$ does not depend on the azimuthal angle $\varphi$. We note that no regularization similar to subtracting the term $1/2T_{nmj}$ in Eq.~(5) of the article is now needed. The sum in Eq.~(\ref{2}) is convergent and, in addition, the regularization used in Eq.~(5) of the article assumes the explicit presence of the coupling constant, which is not the case for the position dependent particle density $n_p({\bf r})$, see Eq.~(\ref{1}). A numerical solution of the self-consistency equation (1) of the article provides us with a set of $\Delta_{\nu}$'s that should be inserted into the above relation for $n_p(\rho,z)$. Our numerical results for $n_p(\rho,z)$ are shown in Fig.~4 of the
article as contour plots. When checking the mean particle density in the center of the trap, one can find that it varies from $4 \times 10^{12}$ to $5\times 10^{12}\,{\rm cm}^{-3}$. So, when taking $n_p = 4.5 \times 10^{12}\,{\rm cm}^{-3}$, we can find $k_F=(3\pi^2 n_p)^{1/3}=5.1\,{\rm \mu m}^{-1}$ and $\lambda_F =2\pi/k_F =1.23\,{\rm \mu m}$. This can be compared with $k_F =\sqrt{\frac{2M}{\hbar^2} (\mu-\hbar\omega_{\perp})} = 3.99\,{\rm \mu m}^{-1}$ and $4.88\,{\rm \mu m}^{-1}$ at $\mu/\hbar\omega_{\perp}=2$ and at $4$, respectively. Both estimates produce close values for $k_F$ in the center of the trap but we obtain different trends for the mean density in the center of the trap. According to the simple estimate, $n_p$ plainly increases with $\mu/\hbar\omega_{\perp}$. Whereas $n_p$ found from the position dependent distribution of the particles slightly oscillates around $4.5 \times 10^{12}\,{\rm cm}^{-3}$.

Having at our disposal possible values of $k_F$ and using our numerical results given in Fig. 1 of the article, we can find $T_c/T_F$ and $k_F\xi_0$ as shown below in Fig.~\ref{fig7} for the simple estimate and Fig.~\ref{fig8} for the estimate based on $n_e(\rho,z)$.

Concluding this section, we would like to note that our results are not sensitive to the regularization procedure used in Eq.~(1) of the article (taken with ${\bf k} \to \{n,m,j\}$) in order to remedy the divergent sum (the ultraviolet divergence). For illustration, Fig.~\ref{fig4} shows numerical results for the critical temperature found when the regularization term $1/(2T_{nmj})$ is abandoned in favor of a simple cut-off $|T_{nmj}-\mu| < \mu\,(\mu \gg \Delta_{nmj}) $, i.e., the sum in Eq.~(1) is only limited to the single-particle states with energies satisfying this inequality. Note that similar cut-off is often used in papers on ultracold atomic gases, see, e.g., Refs.~\onlinecite{bruun,heis,torma}.
\begin{figure}[t]
\resizebox{0.4\columnwidth}{!}{\rotatebox{0}{\includegraphics{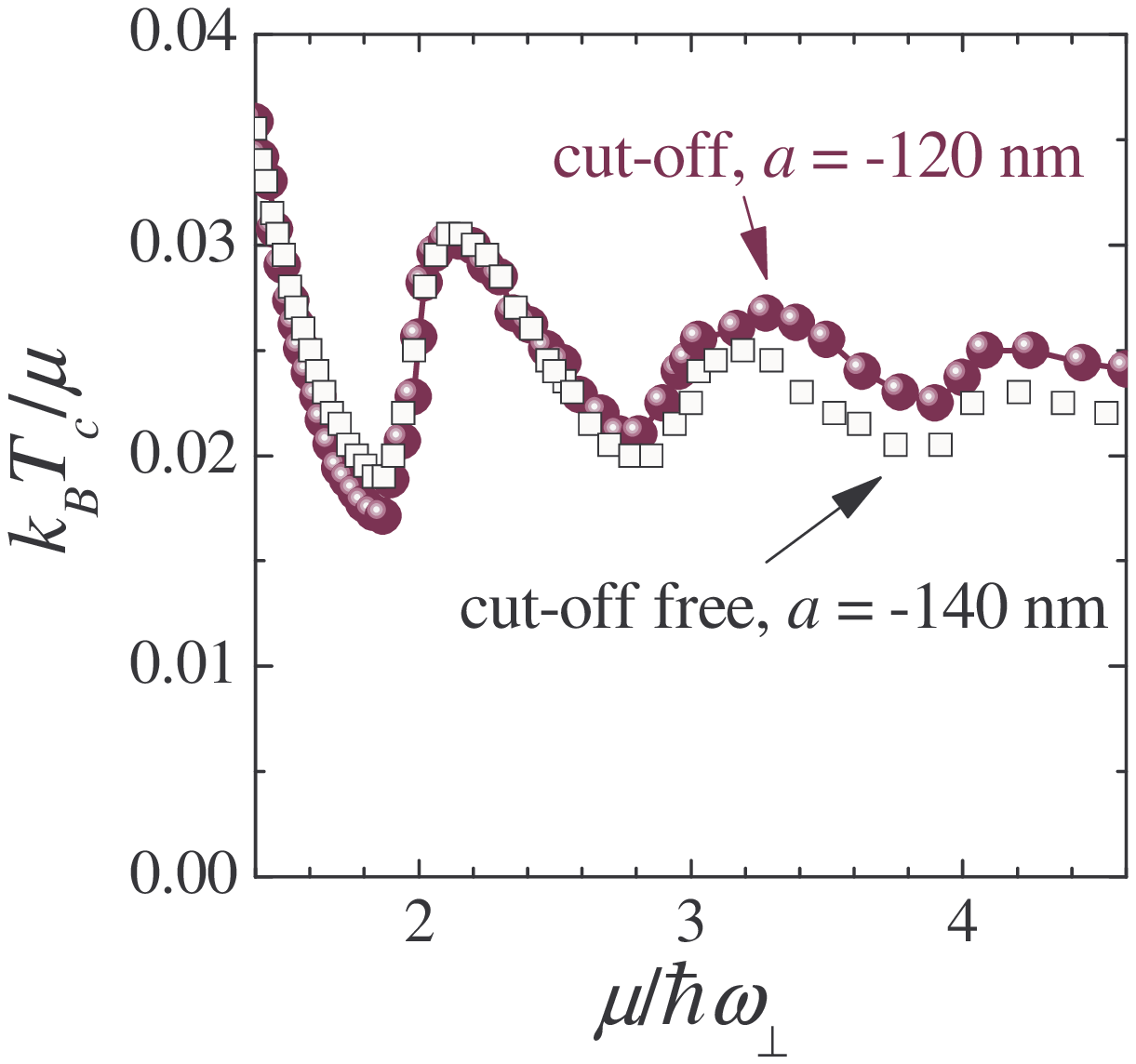}}}
\caption{The critical temperature $T_c$~(in units of $\mu$) versus $\mu/\hbar \omega_{\perp}$ for a cigar-shaped superfluid Fermi gas of $^6Li$: circles are results for $a=-120\,{\rm nm}$ found from the BCS-like self-consistency equation   with the simple cut-off regularization $|T_{nmj} -\mu| < \mu$; squares represent the previous data from Fig. 1(a) of the article.} \label{fig9}
\end{figure}
Though $T_c$ calculated with the cut-off regularization are slightly different as compared to the results given in Fig. 1 of the article, this difference practically disappears when making a small shift in the scattering length, i.e., $a \to a + 20\,{\rm nm}$, see Fig.~\ref{fig9}. Thus, a particular way of ultraviolet regularization does not influence our results on quantum-size oscillations of $T_c$, and the same holds for our conclusions about the size-dependent oscillations in the BCS coherence length.

\section{Superconducting metallic nanowires: parameters used in the calculations}\label{sec_sup2}

Figure \ref{fig5} of the article shows our numerical results calculated for a
cylindrical superconducting aluminum nanowire with the unit cell (periodic boundary conditions) in the $z$ direction $L_z = 50\,{\rm \mu m}$. The coupling constant $g$ is chosen so that $gN(0)=0.18$, with $N(0)$ the bulk density of states at the Fermi level per unit volume and per spin projection (this value of $gN(0)$ is typical of
aluminum, see, e.g., the textbook~\cite{degen}). Effects resulting in the width dependent coupling are beyond the scope of the present investigation (such effects do not influence our main conclusions). The Debye energy is taken the same as in bulk~\cite{degen}, i.e., $\hbar\omega_D/k_B = 375\,{\rm K}$~($\hbar\omega_D = 32.3\,{\rm meV}$). The electron band mass $m_e$ is set to the free-electron
mass in the calculations. The Fermi level is put to $0.9\;{\rm eV}$. We note that this is an effective Fermi level which is used together with the parabolic band approximation to correctly reproduce the period of the quantum-size oscillations. It is a parameter that depends on the interplay between the main crystalline directions and the direction of quantum confinement (for a more detail, see
discussion in Ref.~\onlinecite{shan1}). The above value of the effective Fermi level is justified by a good agreement of theoretical and experimental results for the critical temperature in aluminum nanowires~\cite{shan2}. However, this particular choice is not crucial for our conclusions. We also take into consideration a systematic shift up of the Fermi level with decreasing $d$ for $d < 5-6\,{\rm
nm}$. Such a shift appears if one keeps the same value of the mean electron density when changing $d$. 

\subsection{Surface imperfections}

Our consideration of the superconducting nanowires is in the clean limit and assumes mirror reflections from the boundaries. So the question arises to what extent imperfections of real metallic nanowires can influence our results. We focus on recently fabricated {\itshape high-quality} superconducting nanowires, where the superconducting state survives down to diameters of about $8-10 \,{\rm nm}$ without any signatures of suppression of the critical temperature driven by disorder~[see refs.~\cite{zgi,alt} about ${\rm Al}$ nanowires made of strongly coupled grains and paper~\cite{tian} about ${\rm Sn}$ single-crystalline nanowires). Unlike strongly disordered nanosized superconductors, such high-quality superconducting nanowires exhibit a systematic shift-up of the critical temperature with a reduction in their cross-section ($50\%$ in ${\rm Al}$ specimens and $10-20\%$ in ${\rm
Sn}$ nanowires). As recently shown~\cite{shan2}, it is the quantization of the transverse electron spectrum that shifts $T_c$ up in such nanowires through quantum-size superconducting resonances. Thus, disorder is relatively minor in high-quality superconducting nanowires and the effects of the formation of well
distinguished single-electron subbands become of importance.

To go in a more detail about a role of imperfections in quantum superconducting nanowires, we remark that the surface roughness is here a major disorder mechanism. Indeed, in most papers~\cite{zgi,alt} the mean free path is estimated to approximately follow the nanowire width $\ell \sim d$, i.e., elastic
scattering on the boundary imperfections controls the electron mean free path. So, we need to clarify whether or not the size-dependent drops of the longitudinal BCS coherence length found in the clean limit will be significantly influenced by the surface roughness. The first thing to do with this is to compare the Cooper-pair size $\xi_0$ with the electron mean free path. Beyond a resonant point
$\xi_0$ increases up to microns and, so, we are in the dirty limit, i.e., $\xi_0 \gg \ell$. However, when approaching a point of the quantum-size superconducting resonance, $\xi_0$ drops down to values close to the nanowire diameter, see, e.g., Fig.~\ref{fig5} in the article (this is true for nanowires with $d < 10-20\,{\rm nm}$, for larger diameters the effect is washed out). As the mean free electron path $\ell$ is about $d$, then we obtain $\xi_0 \approx \ell$. So, the
effect of disorder on the superconducting correlations is not significant: the suppression factor in Green's functions is $e^{-r/2\ell} \sim 1$ for $r \sim \xi_0$. Thus, the elastic scattering on the surface imperfections will not significantly alter our conclusions: at points of the size-dependent drops we are close to the clean limit.

The above reasoning assumes that the density of single-electron states at the Fermi level does not change significantly due to surface imperfections. So, the next step is to clarify if the single-electron spectrum is seriously influenced by surface
imperfections. A size-dependent drop of the longitudinal fermionic-pair size occurs due to a subband (subbands) whose bottom (bottoms) is situated in the vicinity of the Fermi surface (see the article). For such a subband the longitudinal motion of electrons is suppressed and the most contribution to the single-electron energy
comes from the transverse spectrum of electrons. This is exactly the reason for the quantum-size driven BCS-BEC crossover. Another consequence of this redistribution of the kinetic energy between the longitudinal and transverse degrees of freedom is long longitudinal wavelengths of the electrons in a resonant subband. The longitudinal electron energy in a subband whose bottom is situated in the Debye
window is about or smaller than the Debye energy. This makes it possible to find that the typical longitudinal wavenumber is smaller than $k < \sqrt{2m_e\omega_D/\hbar}$. So, this simple estimations suggests that the typical longitudinal wavelength of electrons in a resonant subband is larger than $10\,{\rm nm}$. This is significantly larger than the characteristic size of the surface
imperfections usually estimated as being of the order of dimensions of crystalline unit cell (smaller than $1\,{\rm nm}$). Hence, the longitudinal part of the electron spectrum in a resonant subband is stable against surface imperfections. This is not the case for subbands with bottoms positioned far below the Fermi level. Here the electron spectrum can be rather sensitive to the surface roughness
and the corresponding density of states can even be suppressed at the Fermi level in the presence of strong disorder. However, the contribution of such single-electron subbands at a resonant point is only of secondary importance: the major effect comes from the resonant subbands.

In addition to the influence of the surface roughness on the longitudinal electron motion, one should also keep in mind an uncertainty in the transverse energy due to fluctuations in the nanowire diameter. This uncertainty is able to affect the formation of superconducting resonances when it approaches the Debye energy
that controls the selection of the single-electron states making a contribution to the basic superconducting characteristics. The energy spacing between single-electron subbands can be estimated as $\frac{\hbar^2}{2m_e} \frac{\pi^2}{d^2}$. Let us take $d = d_0 + \Delta d$, where $\Delta d$ represents a fluctuation part with values from $0$ to $1\,{\rm nm}$~(this a typical scale for the diameter fluctuations in nanowires with width less than $10$-$20\,{\rm nm}$). The corresponding fluctuation of the inter-subband energy spacing is estimated as $\frac{\hbar^2}{m_e} \frac{\pi^2}{d_0^3}\Delta d$, which can be considered as an
uncertainty in the transverse electron energy. For the narrowest high-quality nanowires with $d=8-10\,{\rm nm}$, such an uncertainty is about $1\,{\rm meV}$, which is much smaller than the Debye energy for aluminum $\hbar\omega_D \approx 32 \,{\rm meV}$. Even for extremely small diameters about $4\,{\rm nm}$ chosen for a simple illustration in the article (to avoid a discussion of many
single-particle subbands responsible for the formation of superconducting resonance at nanowire diameters $\sim 10\,{\rm nm}$), this uncertainty is close to $10\,{\rm meV}$ (when taking $\Delta d =1\,{\rm nm}$). For aluminum this is still three times smaller than the Debye energy.

Another reason for broadening of the single-electron levels is the hybridization with electrons of a semiconductor substrate, which is strongly dependent on the fabrication conditions. However, such a hybridization can be expected to be of importance for specimens with width down to a few monolayers like in ultrathin superconducting nanofilms.

Thus, based on the above discussion we can conclude that our results are quite stable to imperfections of real superconducting nanowires. These imperfections will definitely smooth quantum-size oscillations of $\xi_0$ but will hardly avoid the effect of interest. 

\subsection{Fluctuations}

We work in the mean-field approximation and, so, one more point to discuss is the effect of fluctuations. It is well-known that fluctuations generally play an important role in low-dimensional systems. In superconducting nanowires the main focus is usually on the phase fluctuations of the pair condensate: thermally-activated and quantum phase slips, see, e.g., the paper~\cite{zgi,kost}. These fluctuations lead to a residual resistance remaining below $T_c$ in
narrow nanowires, corrupting the superconducting state. Below we argue that the BCS-BEC crossover driven by quantum confinement is still pronounced above the nanowire diameters, where phase fluctuations proliferate.

Effect of thermal fluctuations is usually estimated through the Ginzburg-Levanyuk parameter calculated from the conventional Ginzburg-Landau functional. However, for {\it quantum nanowires} of width $5-20\,{\rm nm}$ the conventional Ginzburg-Landau formalism can not be applied due to the breakdown of the translational
invariance in the direction perpendicular to the nanowire. A consequence of such a breakdown is that the order parameter strongly varies in the direction perpendicular to the nanowire (see, e.g., Ref.~\onlinecite{shan}). The characteristic length for its spatial variations in this direction is about the nanowire width $d$, which prevents us from using the conventional Ginzburg-Landau formalism which assumes that the order parameter varies slowly on the scale governed by the zero-temperature coherence length. In addition, the 1D Ginzburg-Levanyuk parameter is not relevant either because a superconducting quantum nanowire is an essentially 3D system with multiple single-electron subbands (more than $10-20$ subbands even for $d=4\,{\rm nm}$). Such a multichannel system can not be reduced to effectively one-dimensional case, and this is also seen from the fact that the order parameter has a nontrivial transverse profile that changes significantly with a change of $d$. To overcome this problem, one should abandon the conventional
Ginzburg-Landau formalism in favor of its multichannel version. However, a significant complication here is to accurately incorporate issues related to the quantum-size driven BCS-BEC crossover (e.g., fluctuating Cooper pairs with a nonzero center-of-mass momentum). A simpler option is to rely upon available
experimental results. They suggest that the Ginzburg-Levanyuk parameter is about $0.1-0.15$ in superconducting nanowires with diameters $\approx 10\,{\rm nm}$. In particular, Fig. 1 of the paper~\cite{alt} (for zero magnetic field) demonstrates that the nanowire resistance falls by two orders of magnitude when temperature reduces from $T=T_c$ to $T = 0.85\,T_c$. This is still a rather sharp transition with the thermal broadening of about $\delta T/T_c =0.1-0.15$ and, so, the mean-field treatment looks quite reasonable.

From the same Fig. 1 in the paper~\cite{alt} one can learn that quantum phase fluctuations in an aluminum nanowire with width $10\,{\rm nm}$ result in a residual resistance even below $T/T_c =0.5-0.6$. At this nanowire width such a resistance is smaller than $10^{-4}$ in units of the normal resistance. It is however expected
that for diameters $\lesssim 10\,{\rm nm}$ quantum-phase slips proliferate, which leads to a superconductor-to-normal crossover at $d = d_c$, with $d_c \lesssim 10\,{\rm nm}$~(see, e.g., papers~\cite{zgi,kost}). Recent results of the paper~\cite{zgi,kost} suggest that $d_c\approx 8\,{\rm nm}$. Yet, it is rather difficult to analyze experimental data for very narrow nanowires because it is not possible to completely rule out weak links as the sources of the residual resistance. Thus, we can conclude that the best regime to probe any signatures of the BCS-BEC crossover induced by quantum confinement is to investigate the superconducting nanowires with diameters just above $d_c$. We would ,like to note that the effect of interest is pronounced for $d < 10\,{\rm nm}$ and completely washed out only when $d > 20\,{\rm nm}$. In addition, the subject of our investigation is the zero-temperature coherence length that is one of important parameters controlling the rate of quantum-phase slips, see, e.g., Ref.~\onlinecite{kost,alt}. This makes it possible to expect that our results can be of relevance even for $d < d_c$.

\end{document}